\documentclass[11pt]{article}

\usepackage{graphicx}
\usepackage{multirow}
\usepackage{amsmath,amssymb,amsfonts}
\usepackage[utf8]{inputenc}

\usepackage{mathrsfs}

\usepackage{xcolor}
\usepackage{textcomp}
\usepackage{manyfoot}
\usepackage{booktabs}
\usepackage{setspace}
\usepackage{algorithm}
\usepackage{algpseudocode}
\usepackage{tikz}
\usepackage[margin=1in]{geometry}
\usepackage{url}
\usepackage[caption=false]{subfig}
\usepackage{listings}
\usepackage{natbib}

\usetikzlibrary{shapes.geometric, arrows.meta, positioning, fit}
\usepackage{listings}
\usepackage{natbib}
 \bibpunct[, ]{(}{)}{,}{a}{}{,}%

\begin{document}

\title{Inverse Control Constrained Optimization of Vessel Speed Decisions Under Environmental Risk: Evidence from Arctic Shipping}

\author{
Mauli Pant
\and
Linda J. Fernandez
\and
Indranil Sahoo
}

\date{\today}
\maketitle
\begin{abstract}Understanding how decision-makers balance operational efficiency with environmental and ecological risks is central to vessel navigation. We model vessel speed as a control variable in a constrained optimization framework in which vessel operators balance multiple competing objectives, including transit efficiency, ice-related navigational risk, and whale-related ecological risk. The underlying risk parameters are estimated using over 14 million Automatic Identification System (AIS) observations from the United States Arctic (2010--2019), together with environmental covariates and spatially explicit whale density estimates. The framework incorporates a nonlinear risk objective, vessel heterogeneity, and regularization to ensure stable and interpretable results.

The inferred trade-offs reveal distinct decision-making patterns across vessel groups and navigational statuses. Vessel types such as Tug Tow and Cargo balance operational speed with environmental and ecological considerations. In contrast, several vessel groups, including Fishing, Passenger, and Unspecified vessels, are strongly influenced by ice-related risk, while Pleasure Craft and Tankers exhibit higher sensitivity to whale-related risk.

Across navigational status categories, similar heterogeneity is observed. The dominant status, under way using engine, displays a clear trade-off, whereas other statuses, such as aground and undefined, are strongly shaped by ice-related constraints. Statuses including restricted maneuverability and engaged in fishing exhibit higher estimated sensitivity to whale-related risk, though with substantial uncertainty.

Sensitivity analysis indicates that increasing whale-related risk weighting produces limited changes in model-implied optimal speed, whereas increasing ice-related risk leads to more consistent reductions. These findings suggest that vessel speed decisions are primarily governed by operational and environmental constraints, with whale-related risk playing a secondary role in shaping observed behavior.
\end{abstract}



%


\section{Introduction}\label{sec1}

Decision making in complex transportation systems often involves balancing multiple, competing objectives that are not directly observable. Specifically, for vessels operating in ice affected marine environments (such as the Arctic), navigation decisions are shaped by multiple factors including fuel consumption, travel time, safety considerations, and environmental impacts \citep{Psaraftis2013, Montewka2015}. Risk factors such as exposure to hazardous environmental conditions or potential interactions with marine wildlife are dynamic and directly influence real time navigation decisions. The Arctic is a rapidly evolving geography, characterized by increasing vessel activity, strong seasonal variability, and dynamic ice conditions that are reshaping navigation patterns \citep{Berkman2022ArcticTraffic}. Unlike conventional maritime regions with established shipping corridors, vessel movement in the Chukchi and Beaufort Seas exhibits substantial heterogeneity. Traffic is shaped by a diverse mix of destination-based travel, trans-Arctic transit, and industrial activities such as resource extraction and logistical support\citep{dawson2018temporal}. Satellite based Automatic Identification System (AIS) observations further highlight substantial growth in ship traffic across national and international waters and vessel types \citep{Berkman2022ArcticTraffic}. Despite these changes, there remains limited empirical, data driven understanding of how risks shape real world navigation. Moreover, Arctic traffic comprises diverse transit types including commercial shipping, fishing, tourism, and service vessels. These may operate under heterogeneous objectives and constraints, requiring decision making at finer spatial and temporal scales than traditional route based optimization frameworks. 

In this study, we focus on risk as a measurable and policy relevant component of this broader decision problem. In particular, vessel speed is a key control variable that governs how vessels respond to environmental constraints and ecological risk, making it central to understanding navigation decisions. These decisions are shaped by heterogeneous conditions, including sea ice, weather, and the spatial distribution of marine species, as well as by vessel specific characteristics and implicit risk tolerances. However, the underlying structure governing these trade offs is not directly observed but can be estimated in our data driven analysis.

The increasing availability of high resolution Automatic Identification System (AIS) data, combined with environmental and ecological datasets, provides a unique opportunity to study vessel operator decision making at scale. Panel data linking vessel observations with sea ice concentration, wind conditions, and marine mammal presence enable a detailed characterization of the decision environment. While prior studies have used AIS data to model vessel movement patterns, predict speeds, or detect anomalies \citep{pallotta2013vessel, xing2023fishing, PANT2026125763}, these approaches are primarily descriptive and do not explicitly recover the trade offs underlying observed data.

In addition, existing frameworks typically treat vessel transit and ecological risk as separate modeling problems \citep{pallotta2013vessel, Redfern2013, Silber2014}. As a result, they do not explain how vessels respond to ecological risk, nor how competing objectives such as maintaining speed, avoiding hazardous ice conditions, and minimizing disturbance to marine life are jointly balanced in practice. 

Inverse control constrained optimization provides a principled framework for recovering latent risk based objective functions from observed data. In contrast to standard optimization, which assumes a known objective function to optimize for optimal choices, inverse optimization starts with observed data on the chosen control rule and the constraints to enter into a statistical estimation to recover parameter values of the optimization \citep{ahuja2001inverse}. Early work focused on recovering costs in combinatorial and network settings, such as shortest path and flow problems \citep{ahuja2001inverse, burton1992inverse, zhang1996inverse}.

In this study, we build on recent developments in statistical inverse optimization that explicitly account for noisy observations and imperfect optimality \citep{aswani2018inverse}. These approaches recognize that observed decisions may deviate from exact optima due to incomplete information, heterogeneous constraints, and measurement error. They incorporate regularization and sub optimality loss functions and provide a tractable framework for estimating latent risk parameters from empirical data. While inverse control constrained optimization is often framed in terms of cost minimization \citep{ahuja2001inverse, aswani2018inverse, bertsimas2015data}, we focus on a risk based formulation, where the objective function represents a composite measure of operational, environmental, and ecological risk, corresponding to observable components of the broader decision problem. This formulation is particularly well suited to our setting, where vessel speeds reflect adaptive responses to complex and uncertain environmental conditions rather than exact solutions to a known optimization problem.

This perspective is closely related to inverse control and inverse reinforcement learning (IRL), where the goal is to recover objective or reward functions from observed actions or trajectories \citep{kalman1964when, ng2000algorithms, abbeel2004apprenticeship, ziebart2008maximum}. While these approaches are typically formulated for sequential decision making, they share the central idea that observed data provides indirect information about underlying preferences. In contrast to much of this literature, we focus on local control choices, modeling vessel speed as a continuously observed decision variable. This allows us to infer trade offs at fine spatial and temporal scales without requiring a predefined routing structure.

This formulation is particularly well suited for Automatic Identification System (AIS) data in the United States Arctic region, where vessel movements are not constrained to fixed shipping lanes and navigation decisions are made dynamically in response to local environmental conditions. Within this framework, speed over ground (SOG) is treated as a locally adaptive control variable chosen by vessel operators to minimize transit risk. The approach combines three key components: (i) a data driven baseline estimate of vessel speed derived from GPBoost, representing the empirical control signal; (ii) spatiotemporal whale intensity surfaces obtained from LGCP based models, used to quantify ecological exposure; and (iii) a parametric decision objective that captures penalties associated with alternative speed choices, including deviation from baseline speed, ice related risk, and whale related risk. The unknown weights on these components are recovered through inverse optimization by aligning observed vessel speeds with those that are optimal under the assumed objective, thereby inferring the implicit trade offs governing vessel behavior.

Our analysis allows for quantification of how vessels respond to environmental variability and ecological risk, and supports evaluation of how these responses may change under alternative policy or environmental scenarios. Overall, this work advances the study of vessel transit by introducing a unified, data driven framework that combines machine learning, spatio temporal statistics, and inverse control constrained optimization to recover latent decision making trade offs from observational data and inform policy interventions.

\section{Data Preparation and Exploration}

The analysis integrates multiple data sources that characterize vessel transit, environmental, and ecological conditions. Empirical baseline vessel speed over ground (SOG) is obtained from a previously developed gradient boosted mixed effects model (GPBoost), which characterizes nonlinear relationships and hierarchical dependencies across vessels, spatial locations, and time \citep{PANT2026125763}. The model serves as a data driven representation of typical vessel operating speed.

Ecological risk is quantified using spatiotemporal whale intensity surfaces derived from log Gaussian Cox process (LGCP) models \citep{pant2025lgcp}. These models provide high resolution estimates of beluga and bowhead whale distribution over space and time. Both components provide the key inputs to the inverse control constrained optimization framework, supplying both the empirical baseline for vessel speed and the environmental and ecological data that define the constraints.

AIS data consists of discrete vessel location records rather than continuous movement trajectories. To approximate vessel movement through time and enable time based exposure weighting, pseudo trajectories were constructed by linking consecutive AIS positions into movement segments. Specifically, consecutive records sharing the same vessel identifier (MMSI), spatial grid cell, and daily time index (\texttt{TimeID}) were ordered chronologically using the AIS timestamp (\texttt{BaseDateTime}). A segment was then defined between successive longitude latitude locations within each grouped sequence. Restricting linkage to records within the same vessel, spatial cell, and time index avoids constructing unrealistic long distance movements across large spatial or temporal gaps.

For each segment, the elapsed time between consecutive AIS records was computed as
\[
\Delta t_n = t_n - t_{n-1},
\]
where \(t_n\) denotes the timestamp of observation \(n\). Consecutive point displacement was calculated using the Haversine great circle distance between successive geographic coordinates:
\[
d_n = 2R \arcsin \left(
\sqrt{
\sin^2\left(\frac{\phi_n-\phi_{n-1}}{2}\right)
+
\cos(\phi_{n-1})\cos(\phi_n)
\sin^2\left(\frac{\lambda_n-\lambda_{n-1}}{2}\right)
}
\right),
\]
where \(R\) is Earth’s radius, \(\phi\) denotes latitude, and \(\lambda\) denotes longitude. These quantities were used to estimate segment level movement distance and implied speed.

For each transit segment, the observed vessel speed is denoted by \(v_n\). In the inverse optimization framework, the decision variable is a candidate speed \(v \in [0,v_{\max}]\), whereas \(v_n\) represents the observed speed associated with observation \(n\). Segment level movement quantities were computed between consecutive AIS locations within each pseudo trajectory and assigned to the ending observation of the segment, corresponding to the current AIS record and its associated environmental covariates. Accordingly, all environmental covariates, whale exposure terms, and risk components associated with a segment were evaluated at the ending AIS location.

When a valid AIS reported Speed Over Ground (SOG) value was available at the ending observation, that value was used directly as \(v_n\). Otherwise, speed was estimated from displacement divided by elapsed time:
\[
v_n = \frac{d_n}{\Delta t_n}.
\]

\begin{figure}[t]
\centering
\includegraphics[width=\textwidth]{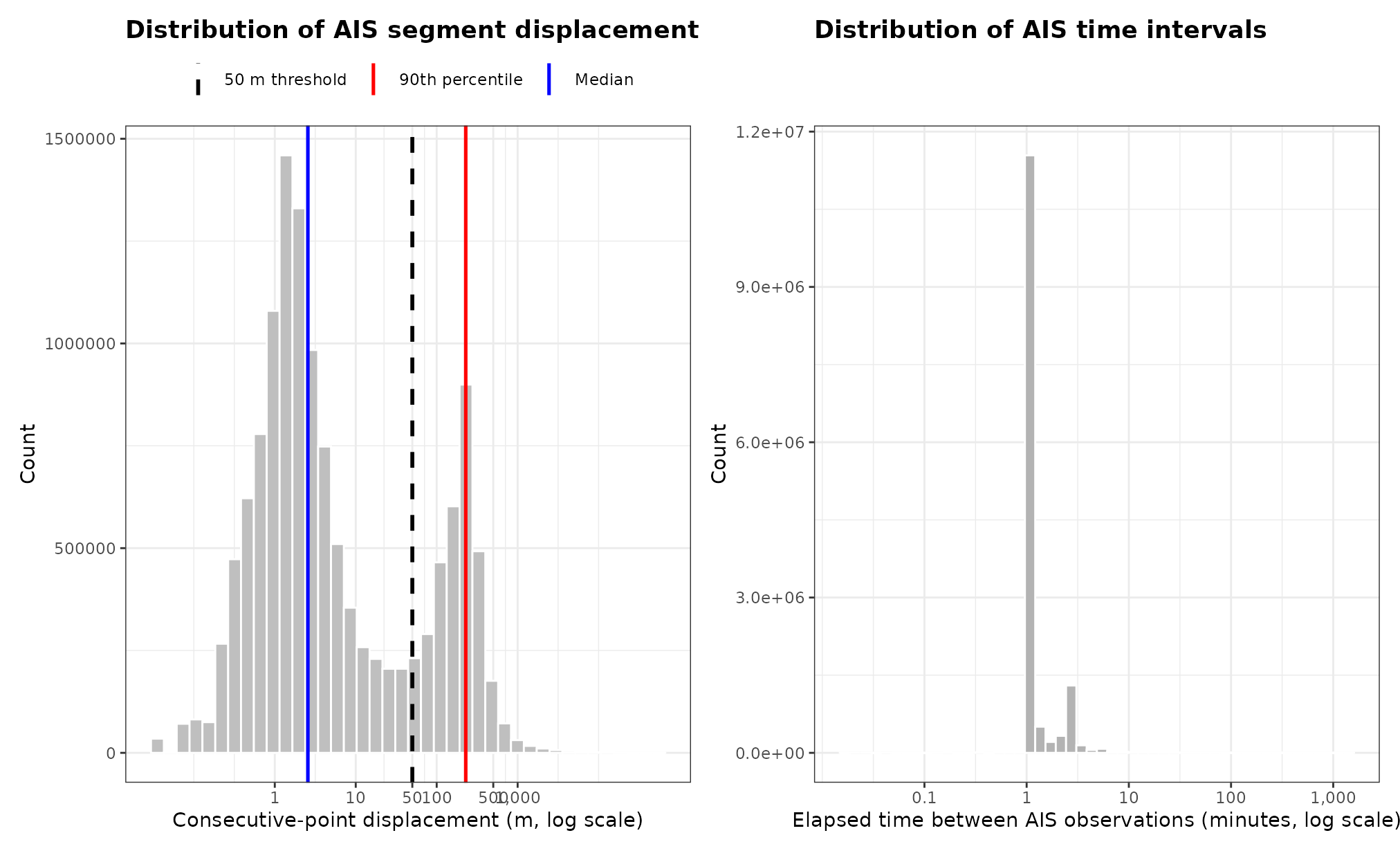}
\caption{
Diagnostic distributions used for pseudo trajectory construction from AIS observations. 
(Left) Distribution of consecutive point displacement between AIS records on a logarithmic scale. 
The blue vertical line denotes the median displacement, the red line denotes the 90th percentile, 
and the dashed black line indicates the imposed 50 m minimum displacement threshold used to reduce GPS positional jitter. 
(Right) Distribution of elapsed time intervals between consecutive AIS observations on a logarithmic scale. 
These diagnostics were used to identify noisy or temporally sparse segments prior to deriving segment level speeds.
}
\label{fig:distance_time_diagnostics}
\end{figure}

Because derived speed estimates depend on both elapsed time and spatial displacement between consecutive AIS records, very small values can generate unstable or artificially large speeds driven by GPS positional jitter rather than true vessel movement. Diagnostic distributions of consecutive point displacement and elapsed time were therefore examined (Figure~\ref{fig:distance_time_diagnostics}).

The distribution of elapsed time between consecutive AIS pings had a median of 0.018 hours (approximately 65 seconds), with the 75th percentile at 0.020 hours and the 99\textsuperscript{th} percentile at 0.10 hours (approximately 6 minutes). These results indicate that most AIS updates occurred at relatively short and regular intervals, although occasional larger temporal gaps were present.

Distance diagnostics likewise showed that most apparent movements were extremely small. Approximately 77.4\% of consecutive point displacements were below 50 m, with a median displacement of approximately 2.2 m and a 75th percentile of 26 m, consistent with GPS positional jitter or highly localized repositioning. In contrast, the upper tail of the displacement distribution contained substantially larger movements, with the 90\textsuperscript{th} percentile exceeding 215 m, indicating genuine vessel transit activity.

Based on these diagnostics, a minimum elapsed time threshold of \(dt_{\min}=0.01\) hours (approximately 36 seconds) and a minimum displacement threshold of 0.05 km (50 m) were imposed before calculating derived segment speeds. Segments not satisfying these conditions were treated as stationary (\(v_n=0\)) to avoid attributing movement to noisy or temporally sparse AIS observations.

\begin{figure}[t]
\centering
\includegraphics[width=0.8\textwidth]{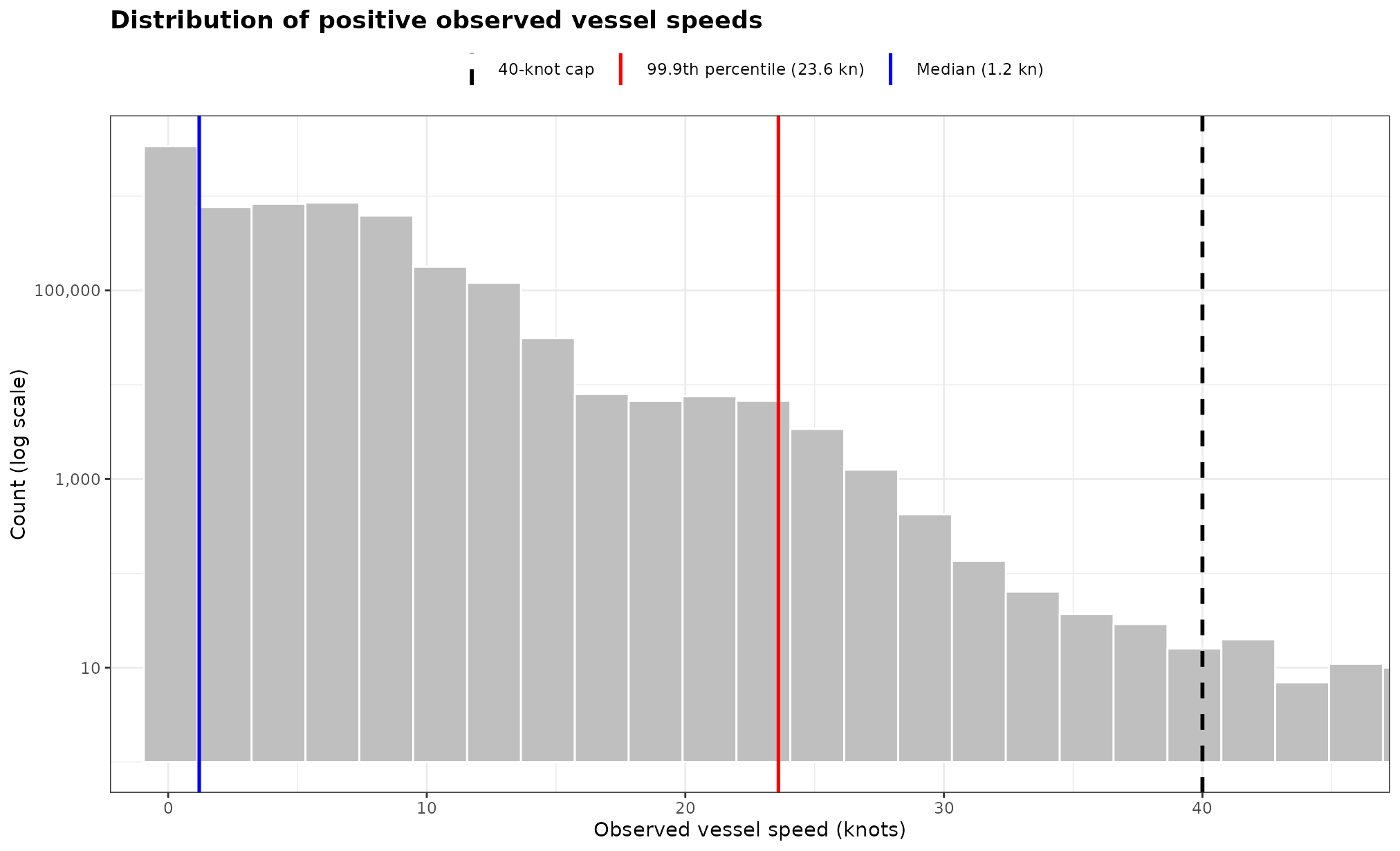}
\caption{
Distribution of positive observed vessel speeds derived from AIS records. 
The blue vertical line denotes the median observed speed, the red line denotes the 99.9th percentile of observed speeds, and the dashed black line indicates the imposed 40 knot upper cap used to exclude implausible outliers. 
The diagnostics show that nearly all observed vessel speeds fall well below the imposed cap, indicating that the threshold functions primarily as a safeguard against extreme AIS artifacts rather than constraining typical operational behavior.
}
\label{fig:speed_diagnostics}
\end{figure}

Figure~\ref{fig:speed_diagnostics} shows the distribution of positive observed vessel speeds derived from AIS records. Observed vessel speeds were capped at 40 knots to remove implausible values. Diagnostic distributions indicate that the 99.9\textsuperscript{th} percentile of observed speeds is approximately 21 knots, with nearly all observations below 30 knots. The imposed cap therefore lies well above the operational range and functions primarily as a safeguard against extreme AIS artifacts rather than constraining typical vessel behavior. For the inverse optimization, candidate speeds were evaluated over a discrete grid spanning 0 to 40 knots with a resolution of 0.5 knots, providing a sufficiently fine approximation to the continuous decision space.

To ensure meaningful exposure calculations, trajectories without positive time intervals were removed. Because segment level exposures are defined in terms of elapsed time, trajectories with zero total duration cannot contribute to the estimation of risk trade offs. These cases typically arise from isolated AIS records or duplicate timestamps and represent less than 0.01\% of the dataset.

Segments with missing baseline speed \(\mu(s_n)\) predictions or invalid environmental covariates were also excluded. These observations constitute only a negligible fraction of the data. Zero SOG segments were retained in the analysis. For observations with \(SOG = 0\), the baseline speed prediction is set to
\[
\mu(s_n) = 0,
\]
so that stationary vessels are treated as consistent with expected behavior and do not incur a deviation penalty in the objective function. This baseline quantity \(\mu(s_n)\) is distinct from the observed segment level speed \(v_n\), which is observed vessel SOG for each segment. At the same time, stationary segments continue to contribute to time based exposure.

Segment level exposure terms such as
\[
\text{Whale\_Time\_Exposure}_n = D(s_n)\Delta t_n
\quad \text{and} \quad 
\text{Ice\_Time\_Exposure}_n = I(s_n)\Delta t_n
\]
remain positive whenever \(\Delta t_n > 0\), even if \(v_n = 0\). Thus, vessels that are stopped or drifting in whale habitat or ice covered waters still accumulate exposure over time.

Finally, segment level risks are weighted by the observed time between AIS messages \((\Delta t_n)\), so that environmental conditions experienced for longer durations contribute proportionally more to the inferred trade offs. This time weighted formulation is consistent with the latent risk framework, where vessel speed is interpreted as a response to cumulative exposure to environmental conditions, including whale presence and ice related navigational constraints. This construction provides the foundation for linking observed vessel speed to inferred navigation trade offs in subsequent analysis.

\subsection{Descriptive Summary of Vessel Activity}

The dataset consists of approximately 14.3 million AIS observations (\(n=1,\dots,N\)) from 523 unique vessels over 1,194 days. Vessel activity is highly concentrated among a small number of vessel groups. Tug and tow vessels dominate the dataset, accounting for approximately 70.1\% of all observations, followed by cargo vessels (13.7\%) and other vessel types (7.8\%). Passenger and fishing vessels contribute smaller shares of 3.0\% and 2.4\%, respectively, while all remaining vessel groups each account for less than 2\% of observations.

AIS navigational status is similarly dominated by vessels \textit{under way using engine}, which account for 77.3\% of observations. Other common statuses include \textit{at anchor} (6.7\%), \textit{undefined} (5.1\%), \textit{restricted maneuverability} (3.8\%), and \textit{moored} (3.2\%). All remaining statuses occur relatively infrequently, each representing less than 3\% of the dataset.

Observed vessel speeds \(v_n\) are generally low, with a mean of 1.44 knots and a median of 0 knots, reflecting a substantial proportion of stationary or near stationary observations. The 90th percentile of observed speed is 6.2 knots, and the 99th percentile is 12 knots, indicating that most vessels operate within a relatively moderate speed range, with only a small fraction of high speed movements.

Environmental conditions in the study region are characterized by high ice concentration \(I(s_n)\), with a mean value of 0.81 and a standard deviation of 0.13. This indicates that vessels frequently operate in ice covered environments, reinforcing the importance of environmental constraints in shaping navigation behavior.
\section{Inverse Control Constrained Optimization Framework}

AIS observations are treated as indexed records $n = 1,\dots,N$, where $N$ denotes the total number of observations collected during the study period (July–October, 2010–2019). For each record $n$, we define the observed state as
\begin{equation}
\label{eq:state}
s_n = (i_n, c_n, t_n, \mathbf{x}_n),
\end{equation}
where $i_n$ is the vessel identifier (MMSI), $c_n$ is a spatial grid cell, $t_n$ is a time index, and $\mathbf{x}_n$ is a vector of environmental and operational covariates, including sea ice concentration, wind, bathymetry, distance to coast, course over ground (COG), whale intensity, vessel group, and navigational status. The observed control variable is vessel speed over ground (SOG),
\begin{equation}
\label{eq:speed}
v_n \in \mathbb{R}_{\ge 0}.
\end{equation}
We retain observations with $v_n = 0$, as stationary or slow moving vessels can still experience prolonged exposure (e.g., lingering within whale habitat or ice).
Within the inverse optimization framework, the observed segment speed \(v_n\) is interpreted as the realized outcome of a latent vessel decision process under environmental and operational constraints. The framework assumes that observed speeds arise from approximately minimizing an unobserved risk objective that balances deviation from empirically typical operating speeds against competing environmental and ecological risks. Accordingly, the inverse problem consists of recovering the latent trade off structure that best explains the observed AIS speed decisions. Equation~\eqref{eq:objective} formalizes this inverse optimization problem by estimating the trade off parameters that minimize the aggregate suboptimality gap between observed vessel speeds and the speeds implied by the latent risk objective.
\subsection{Inverse Control Objective Specification}

We model observed vessel speed choices as arising from optimizing an objective function, with navigation risk divided into three components: deviation from an empirical baseline speed (Section~3.2), whale related risk (Section~3.3), and ice related risk (Section~3.4).

Let $\mu(s_n)$ denote the empirical baseline speed estimated by the first stage predictive model. For each observation $n$, let $v_n$ denote the observed segment level speed, while $v \in [0, v_{\max}]$ denotes a candidate speed in the decision space.

For a candidate speed $v$ and observation specific state $s_n$, the risk objective is
\begin{equation}
\label{eq:objective_risk}
R_n(v;s_n,g_n,\boldsymbol{\theta})=
\Delta t_n \left[
\frac{\tfrac{1}{2}(v - \mu(s_n))^2}{c_{\Delta}}
+
\theta_{W,g_n} \,\frac{D(s_n)\left(v+ v^m\right)}{c_W}
+
\theta_{I,g_n} \,\frac{I(s_n)\,v^2 + \bigl(v - v_{\mathrm{safe}}(s_n)\bigr)_+^2}{c_I}
\right],
\end{equation}
where \(g_n\) denotes the vessel group or navigational status category associated with observation \(n\), and \(\theta_{W,g_n}\) and \(\theta_{I,g_n}\) are the corresponding group-specific weights. Here, \((x)_+ = \max(x,0)\) denotes the positive part operator. Thus, \(\bigl(v - v_{\mathrm{safe}}(s_n)\bigr)_+^2\) penalizes only speeds exceeding the local safe-speed threshold. Speeds below or equal to \(v_{\mathrm{safe}}(s_n)\) receive no exceedance penalty. \(D(s_n)\) denotes whale encounter intensity, \(I(s_n)\) denotes ice concentration, and \(v_{\mathrm{safe}}(s_n)\) is an ice-concentration-dependent safe-speed benchmark \citep{HSVA2014FuelConsumption}. The exponent \(m > 0\) governs the nonlinear scaling of whale-related acoustic risk with vessel speed.

The estimated weights satisfy
\begin{equation}
\label{eq:simplex}
\theta_{W,g} + \theta_{I,g} = 1,
\qquad
\theta_{W,g},\theta_{I,g} \ge 0,
\quad \forall g.
\end{equation}

The first term in Equation \eqref{eq:objective_risk} is a fixed quadratic penalty for deviation from the empirical baseline speed. Since $\mu(s_n)$ is estimated from the predictive model, it represents the typical speed implied by prevailing environmental and operational conditions. This term therefore anchors the inverse control objective around observed operating behavior without introducing an additional trade off parameter.

The second term in Equation \eqref{eq:objective_risk} allows for nonlinear acoustic effects through a flexible power specification. Expressing the exponent as $m$ enables direct comparison with acoustic scaling relationships, where approximately linear scaling corresponds to $m \approx 1$. The ice related term combines a continuous quadratic operating risk, $I(s_n)\,v^2$, with an additional quadratic exceedance penalty, $\bigl(v - v_{\mathrm{safe}}(s_n)\bigr)_+^2$, so that ice affects behavior both through general navigation difficulty and through speeds above the safe speed benchmark.

To improve comparability across components, each term is normalized by an empirical scale factor $c_{\Delta}$, $c_W$, and $c_I$, computed from the corresponding observed term distribution using the 95th percentile. This scaling reduces numerical dominance due to differences in units or magnitude and allows the inferred whale versus ice trade off to be interpreted more directly. We now describe each component of the risk objective function in Equation~\eqref{eq:objective_risk} in detail.

\subsection{Empirical Baseline Speed Model}

To characterize the transit control variable, vessel SOG, under observed environmental and navigational conditions, we employ a statistical speed model based on the GPBoost framework developed in \cite{PANT2026125763}. GPBoost combines gradient boosting decision trees with grouped random effects, enabling flexible nonlinear modeling while accounting for structured heterogeneity arising from vessel identity, vessel location, time, vessel group, and navigation status.

For each state \(s_n\), defined by environmental covariates (sea ice concentration, wind conditions, bathymetry, and course over ground (COG)) and vessel characteristics (vessel group and navigational status), the GPBoost baseline model yields a predicted mean vessel speed conditional on \(SOG>0\), denoted by \(\mu(s_n)\). We interpret \(\mu(s_n)\) as the empirical baseline speed, representing the typical vessel speed expected under similar environmental and operational conditions.

Deviations from this empirical baseline are penalized through the quadratic term
\begin{equation}
\label{eq:baseline_term}
\frac{\tfrac{1}{2}(v - \mu(s_n))^2}{c_{\Delta}},
\end{equation}
where \(v\) denotes the candidate decision speed and \(c_{\Delta}\) is a scaling constant. This component encourages candidate speeds to remain close to historically observed operating behavior while still allowing trade offs with whale related and ice related risk components.

This baseline captures systematic heterogeneity in vessel speed across vessel groups and navigational status, as well as spatial and temporal variation induced by environmental conditions. Deviations from $\mu(s_n)$ therefore reflect departures from typical operating tendencies. Within the inverse optimization framework, deviations from the empirical baseline enter as a quadratic anchoring term, which centers the inferred decision problem around empirically typical speeds for a variety of vessel groups. 

The scaling factor \(c_{\Delta}\) normalizes the quadratic baseline deviation term so that its magnitude is comparable to the whale related and ice related risk components in the objective function. Specifically, \(c_{\Delta}\) was defined empirically from the distribution of observed squared deviations between AIS reported vessel speeds and the GPBoost baseline predictions, $(v_n - \mu(s_n))^2$. To reduce sensitivity to extreme outliers while preserving the operational scale of typical deviations, \(c_{\Delta}\) was set equal to the 95\textsuperscript{th} percentile of this empirical distribution. Consequently, deviations commonly observed in the AIS data contribute on a scale comparable to the environmental risk terms, while unusually large deviations receive proportionally larger penalties.

The empirical baseline speed \(\mu(s_n)\) was obtained from the positive speed component of the two stage GPBoost framework. Conditional on \(SOG>0\), vessel speed was modeled as
\begin{equation}
\label{eq:gpboost_baseline}
\sqrt{SOG_n}
=
f(s_n)
+
b_{\mathrm{MMSI}(n)}
+
b_{\mathrm{cell}(n)}
+
b_{\mathrm{time}(n)}
+
\varepsilon_n,
\end{equation}
where \(f(s_n)\) denotes the boosted tree fixed effect component based on environmental and operational covariates, \(b_{\mathrm{MMSI}(n)}\), \(b_{\mathrm{cell}(n)}\), and \(b_{\mathrm{time}(n)}\) denote vessel specific, spatial, and temporal random effects, respectively, and \(\varepsilon_n\) is the residual error term. The empirical baseline speed \(\mu(s_n)\) was then obtained by transforming predictions back to the original speed scale.

To avoid overfitting and ensure that the empirical baseline remained unbiased, \(\mu(s_n)\) was constructed using out of fold predictions from a leave one year out (LOYO) cross validation scheme. Specifically, for each held out year, the GPBoost model was trained using all remaining years and then used to predict vessel speeds for the excluded year. This procedure ensured that baseline predictions for each observation were generated without using information from the same temporal fold, thereby preserving the integrity of the downstream inverse optimization inference.

Next, we construct environmental risk components to represent spatial and temporal variation in ecological exposure and operational constraints.

\subsection{Whale Encounter Intensity}

Spatial and temporal variation in whale sightings is represented using predictions from a spatiotemporal Log Gaussian Cox Process (LGCP) fitted to aerial survey observations \citep{pant2025lgcp}. This model produces a continuous, nonnegative encounter intensity surface, which represents the expected whale encounter intensity at state $s_n$, conditional on environmental covariates and spatiotemporal structure. Predictions are generated on a spatial grid and mapped to vessel locations via nearest neighbor assignment, yielding $D(s_n)$ for each AIS observation.

Whale related risk is modeled as
\begin{equation}
\label{eq:whale_risk}
\frac{D(s_n)\,\bigl(v + v^{m}\bigr)}{c_W},
\end{equation}
where $m>0$ governs the nonlinear scaling of acoustic risk with vessel speed. The linear component $D(s_n)v$ captures speed weighted interaction exposure, while the nonlinear term $D(s_n)v^{m}$ represents acoustic disturbance, allowing risk to scale superlinearly with vessel speed when $m>1$. Under this parameterization, $m=1$ corresponds to approximately linear scaling of acoustic exposure, while larger values induce increasingly convex responses, thereby assigning greater weight to high speed vessel operations. Empirical and theoretical studies suggest that radiated noise from ships increases nonlinearly with vessel speed, often approximated by power law relationships with exponents in the range of 2–4 \citep{Ross1976, McKenna2012, Erbe2012}. Analogous to \(c_{\Delta}\), the scaling factor \(c_W\) was defined as the 95\textsuperscript{th} percentile of the empirical distribution of
$D(s_n)\bigl(v_n + v_n^{m}\bigr)$, evaluated using observed segment level speeds \(v_n\) and whale density estimates \(D(s_n)\).

\subsection{Ice Conditions and Safe Speed Guidance}

Sea ice concentration at state $s_n$ is denoted by $I(s_n)\in[0,10]$ in tenths. To represent operational guidance under ice conditions, we use a piecewise linear approximation to the \cite{HSVA2014FuelConsumption} report safe speed recommendations:
\begin{equation}
\label{eq:vsafe}
v_{\mathrm{safe}}(I)=
\begin{cases}
19 - \dfrac{14}{5}I, & 0 \le I \le 5,\\[6pt]
5 - \dfrac{1}{5}(I-5), & 5 < I \le 10.
\end{cases}
\end{equation}
Ice related risk is then modeled as
\begin{equation}
\label{eq:ice_risk}
\frac{I(s_n)\,v^2 + \bigl(v - v_{\mathrm{safe}}(s_n)\bigr)_+^2}{c_I}
\end{equation}
where $(x)_+ = \max(x,0)$ and $v_{\mathrm{safe}}(s_n)$ denotes the ice dependent safe speed benchmark defined in Equation~\eqref{eq:vsafe}.Combining these components, vessel speed choice is represented as minimizing a latent risk function composed of (i) deviation from an empirical baseline and (ii) a group-specific trade off between whale-related and ice-related risks. The relative importance of whale versus ice risk is governed by the estimated group-specific weights, while the baseline deviation term serves as a fixed anchoring component.

The first term captures increasing navigation difficulty as ice concentration rises, while the exceedance term penalizes speeds above the guidance based safe speed benchmark. The quadratic structure reflects the nonlinear escalation of navigational risk at higher speeds in ice covered waters. Analogous to \(c_{\Delta}\) and \(c_W\), the scaling factor \(c_I\) is defined as the 95\textsuperscript{th} percentile of the empirical distribution of the ice related risk term, evaluated using observed segment level speeds and environmental conditions.

Combining these components, vessel speed choice is represented as minimizing a latent risk function composed of (i) deviation from an empirical baseline and (ii) a trade off between whale related and ice related risks. The relative importance of whale versus ice risk is governed by the estimated weights, while the baseline deviation term serves as a fixed anchoring component.


The functional forms used in the inverse control objective were selected to balance interpretability, numerical stability, and alignment with domain knowledge. The quadratic baseline deviation term provides a smooth anchoring around the empirical baseline speed. The whale related term allows for nonlinear acoustic effects through a flexible power specification, consistent with empirical and theoretical studies showing that ship radiated noise increases nonlinearly with vessel speed, often following power law relationships with exponents in the range of 2–4 \citep{Ross1976, McKenna2012, Erbe2012}. The quadratic ice related term captures both continuous navigation difficulty and nonlinear exceedance of safe speed guidance.
\subsection{Inverse Control Constrained Optimization Framework}

We model vessel speed choice as an optimal control decision under incomplete information and operational constraints. For each state $s_n$, the chosen segment speed $v_n$ is assumed to minimize a latent objective risk function $R_n(v; s_n, \boldsymbol{\theta})$ over the feasible set
\begin{equation}
\label{eq:feasible_set}
\mathcal{A}(s_n) = \{ v \in \mathbb{R}_{\ge 0} : 0 \le v \le v_{\max} \}.
\end{equation}
For given parameters, the optimal speed is defined as
\begin{equation}
\label{eq:optimal_control}
v^\star(s_n;\boldsymbol{\theta})
=
\arg\min_{v \in \mathcal{A}(s_n)} \mathbb{E}\big[ R_n(v; s_n, \boldsymbol{\theta}) \big],
\end{equation}
where the expectation represents uncertainty in the latent decision environment. In practice, this expectation is approximated using observed and predicted mean quantities (e.g., $\mu(s_n)$ and $D(s_n)$), yielding a tractable plug in representation of the decision problem. Accordingly, the optimization is implemented using the empirical segment level objective function \(R_n(v;s_n,\boldsymbol{\theta})\), constructed from observed and predicted quantities.

Observed speeds are treated as approximately optimal, allowing for deviations due to unobserved constraints, measurement noise, and behavioral variability. We quantify this using the suboptimality gap
\begin{equation}
\label{eq:gap}
\Delta_n(\boldsymbol{\theta})
=
R_n(v_n; s_n, g_n, \boldsymbol{\theta})
- \min_{v \in \mathcal{A}(s_n)} R_n(v; s_n, g_n, \boldsymbol{\theta}).
\end{equation}

We estimate the latent trade off parameters by minimizing the total gap:
\begin{equation}
\label{eq:objective}
\hat{\boldsymbol{\theta}}
=
\arg\min_{\boldsymbol{\theta}}
\sum_{n=1}^N \max\{\Delta_n(\boldsymbol{\theta}), 0\}
+
\lambda \|\boldsymbol{\eta}\|^2,
\end{equation}
where $\boldsymbol{\eta}$ denotes the vector of unconstrained logit parameters governing the whale-ice trade off across groups, and $\lambda$ is a regularization parameter controlling shrinkage of these parameters toward zero.

To allow trade offs to vary across observational units, we introduce group specific structure in the logit parameter:
\begin{equation}
\label{eq:group_eta}
\eta_g = \alpha_0 + \alpha_g,
\end{equation}
with corresponding weights
\begin{equation}
\label{eq:group_theta}
\theta_{W,g} = \frac{1}{1 + e^{\eta_g}}, 
\qquad
\theta_{I,g} = \frac{e^{\eta_g}}{1 + e^{\eta_g}}.
\end{equation} \noindent The empirical observation level objective function is given by
\begin{equation}
\label{eq:risk}
\begin{aligned}
R_n(v;s_n,g_n,\boldsymbol{\theta})
= \Delta t_n \Bigg[
&\frac{\tfrac{1}{2}(v - \mu(s_n))^2}{c_{\Delta}}
+ \theta_{W,g_n} \frac{D(s_n)\,\bigl(v + v^m\bigr)}{c_W} \\
&\quad
+ \theta_{I,g_n} \frac{I(s_n)\,v^2
+ \bigl(v - v_{\mathrm{safe}}(s_n)\bigr)_+^2}{c_I}
\Bigg],
\qquad v \in [0, v_{\max}].
\end{aligned}
\end{equation} \noindent Here, the first term is a fixed baseline deviation component, and
\(\theta_{W,g_n}\) and \(\theta_{I,g_n}\) govern the trade off between
whale related and ice related risks.

The optimization problem in Equation~\eqref{eq:objective} is solved numerically using the derivative-free Bound Optimization by Quadratic Approximation (BOBYQA) algorithm \citep{Powell2009}, implemented through the \texttt{nloptr} package in \texttt{R}. The BOBYQA algorithm was selected because the empirical objective function is nonlinear, potentially nonconvex, and contains piecewise components arising from the positive part exceedance penalty and the discrete approximation of the admissible speed space. These features make analytical gradients difficult to obtain reliably and reduce the suitability of standard gradient based optimization methods. BOBYQA provides a stable derivative free trust region approach for bounded nonlinear optimization and performed reliably across the large scale bootstrap estimation procedure. Since the latent risk objective contains nonlinear and piecewise components, including the positive part operator in the ice exceedance penalty, analytical gradients are not readily available. Accordingly, the optimization is performed using direct search over the unconstrained parameter vector $\boldsymbol{\eta}$ governing the whale-ice trade off weights.

For a given parameter vector, the objective function is evaluated by computing the observed segment level risk and comparing it against the minimum achievable risk over a discretized admissible speed grid,
\[
\mathcal{V} = \{0,\texttt{grid\_by},\dots,v_{\max}\},
\]
where candidate speeds are evaluated at increments of 0.5 knots between 0 and 40 knots. For each observation, the minimum attainable risk is approximated by exhaustive evaluation of the latent risk function over this candidate grid, yielding the empirical suboptimality gap in Equation~\eqref{eq:gap}. The optimization then iteratively updates the trade off parameters to minimize the aggregate penalized gap across all observations.

To improve numerical stability and comparability across objective components, the baseline deviation, whale related risk, and ice related risk terms are normalized using empirical scaling constants computed from the 95\textsuperscript{th} percentile of the corresponding observed term distributions. Optimization was initialized using neutral starting values and constrained within bounded parameter ranges to avoid extreme solutions.
Uncertainty in the estimated trade off parameters is assessed via a stratified bootstrap (described in Algorithm \ref{alg:inverse_risk_full}). Within each vessel group or navigational status stratum, pseudo trajectories are sampled with replacement, and all associated observations are retained. This preserves within trajectory 
dependence while approximating the sample size. The model is re-estimated for each replicate, and parameter uncertainty is quantified using the empirical bootstrap distribution.
\begin{algorithm}[!htbp]
\caption{Inverse risk estimation with stratified bootstrap}
\label{alg:inverse_risk_full}
\footnotesize
\begin{algorithmic}[1]
\Require Segment-level dataset $\mathcal{D}$, number of bootstrap replicates $B$, bootstrap cap $N_{\text{boot}}$
\Ensure Group-specific risk weights $\bigl(\widehat{\theta}_{W,g},\widehat{\theta}_{I,g}\bigr)$ and bootstrap uncertainty intervals

\State \textbf{Data preparation:}
\State Filter $\mathcal{D}$ to retain valid segment-level observations with finite observed speed $v_n$, baseline speed $\mu(s_n)$, whale intensity $D(s_n)$, ice concentration $I(s_n)$, safe speed $v_{\mathrm{safe}}(s_n)$, and positive segment duration $\Delta t_n$
\State Assign each observation $n$ to a categorical group index $g_n$ (e.g., vessel group or navigational status)

\State \textbf{Define candidate controls and scaling:}
\State Construct the admissible speed grid $\mathcal{V}=\{0,\texttt{grid\_by},\dots,v_{\max}\}$
\State Compute scaling constants $c_{\Delta}$, $c_W$, and $c_I$ using empirical quantiles of the corresponding objective components

\State \textbf{Parameterization of group-specific weights:}
\State Parameterize group-specific whale-risk weights using unconstrained parameters $\eta_g$:
\[
\theta_{W,g} = \frac{1}{1 + \exp(\eta_g)}
\]
\State Define the corresponding ice-risk weights as
\[
\theta_{I,g} = 1 - \theta_{W,g}
\]
\State Let $\boldsymbol{\eta}=(\eta_1,\dots,\eta_G)$ and $\boldsymbol{\theta}=\{(\theta_{W,g},\theta_{I,g})\}_{g=1}^G$

\State \textbf{Risk evaluation and objective:}
\State Define the latent risk function $R_n(v;s_n,g_n,\boldsymbol{\theta})$ as in Equation~(\ref{eq:risk})
\State For each observation, compute the suboptimality gap
\[
\Delta_n(\boldsymbol{\theta})
=
R_n(v_n;s_n,g_n,\boldsymbol{\theta})
-
\min_{v\in\mathcal{V}} R_n(v;s_n,g_n,\boldsymbol{\theta})
\]
\State Estimate $\widehat{\boldsymbol{\eta}}$ by minimizing the penalized objective in Equation~(\ref{eq:objective})
\State Transform $\widehat{\boldsymbol{\eta}}$ to recover fitted group-specific weights $\bigl(\widehat{\theta}_{W,g},\widehat{\theta}_{I,g}\bigr)$ for all groups $g$

\State \textbf{Bootstrap uncertainty estimation:}
\For{$b=1,\dots,B$}
    \State Draw a stratified bootstrap sample $\mathcal{D}_b$ of approximately $N_{\text{boot}}$ observations from $\mathcal{D}$, stratifying by $g_n$
    \State Preserve the distribution of observations across groups
    \State Reconstruct group indices $g_n^{(b)}$ for observations in $\mathcal{D}_b$
    \State Initialize optimization using the full-dataset estimate $\widehat{\boldsymbol{\eta}}$ as a warm start
    \State Re-estimate the model on $\mathcal{D}_b$
    \State Store bootstrap estimates $\bigl(\widehat{\theta}_{W,g}^{(b)},\widehat{\theta}_{I,g}^{(b)}\bigr)$ for all groups $g$
\EndFor

\State Summarize uncertainty using the empirical bootstrap distribution of $\bigl(\widehat{\theta}_{W,g}^{(b)},\widehat{\theta}_{I,g}^{(b)}\bigr)$
\end{algorithmic}
\end{algorithm}

\section{Results}
\subsection{Selection of the Acoustic Exponent ($m$)}

We assessed sensitivity to the acoustic exponent governing the whale related risk term by estimating the model for 
$m \in \{1,2,3\}$ under identical settings. Prior literature suggests that acoustic exposure scales nonlinearly 
with vessel speed, often following power law relationships with exponents in the range of 2–4 \citep{Ross1976, McKenna2012, Erbe2012}. In our formulation, this corresponds to $m \in \{1,2,3\}$. Model fit was evaluated using the aggregate suboptimality objective, where lower values indicate better alignment between observed speeds and inferred decision behavior. The results show that $m = 2$ provides the best overall fit, while both lower and higher values lead to a deterioration in fit.

In particular, $m = 1$  leads the model to assign nearly all weight to whale related risk, effectively collapsing the trade off and making vessel behavior appear dominated by whale exposure. In contrast, $m = 3$ produces the opposite pattern, with the model assigning nearly all weight to ice related risk. These solutions indicate that overly weak or overly steep acoustic scaling distorts the inferred trade offs and fails to capture meaningful behavioral heterogeneity. In this case, $m = 2$, corresponding to a quadratic relationship between speed and acoustic risk, provides the most interpretable and plausible specification. This choice balances flexibility and stability, preserves heterogeneity across vessel groups and navigational status, and yields the best overall fit to observed behavior.

\subsection{Primary Estimates of Behavioral Trade offs Across Vessel Groups}
\begin{table}[h]
\centering
\caption{Estimated trade off weights between whale related risk ($\widehat{\theta}_W$) and ice related risk ($\widehat{\theta}_I$) by vessel group.}
\label{tab:theta_vg}
\begin{tabular}{lcc}
\toprule
\textbf{Vessel Group} & $\widehat{\theta}_W$ & $\widehat{\theta}_I$ \\
\midrule
Tug Tow         & 0.5193 & 0.4807 \\
Cargo           & 0.2334 & 0.7666 \\
Cruise Ship     & 0.0551 & 0.9449 \\
Dredger         & 0.0085 & 0.9915 \\
Fishing         & 0.0001 & 0.9999 \\
Other           & 0.0965 & 0.9035 \\
Passenger       & 0.0001 & 0.9999 \\
Pilot Vessel    & 0.0523 & 0.9477 \\
Pleasure Craft  & 0.9991 & 0.0009 \\
Reserved        & 0.9789 & 0.0211 \\
SAR             & 0.0026 & 0.9974 \\
Service Ship    & 0.2804 & 0.7196 \\
Tanker          & 0.9957 & 0.0043 \\
Unspecified     & 0.0228 & 0.9772 \\
\bottomrule
\end{tabular}

\end{table}
Table~\ref{tab:theta_vg} reports the estimated trade off weights obtained from the model fit on the full dataset.
These estimates serve as the calibrated parameter values used for interpretation, while uncertainty is assessed through bootstrap resampling.
The full dataset estimates reported in Table~\ref{tab:theta_vg} are broadly consistent with the bootstrap results presented in Table~\ref{tab:vessel_theta}. In most cases, the point estimates from the main fit lie well within the corresponding bootstrap confidence intervals, indicating stability of the inferred trade offs. For vessel groups such as fishing, passenger, tanker, and tug tow, the bootstrap intervals are relatively narrow and centered around the full dataset estimates, suggesting precise and well identified behavioral patterns.

In contrast, several vessel groups including cruise ships, dredgers, SAR, and service vessels exhibit wide bootstrap confidence intervals that span a large portion of the feasible parameter space. Although the full dataset estimates for these groups suggest particular trade offs, the associated uncertainty indicates that these patterns are less precisely identified. Thus, our results demonstrate that while the full dataset model provides reliable point estimates, the degree of uncertainty varies substantially across vessel groups. This highlights the importance of interpreting inferred behavioral trade offs in conjunction with their bootstrap variability.

\begin{table}[ht]
\centering
\caption{Estimated trade off weights between whale related risk ($\widehat{\theta}_W$) and ice related risk ($\widehat{\theta}_I$) by navigational status.}
\label{tab:theta_status}
\begin{tabular}{lcc}
\toprule
\textbf{Status} & $\widehat{\theta}_W$ & $\widehat{\theta}_I$ \\
\midrule
0  & 0.2748 & 0.7252 \\
1  & 0.5940 & 0.4060 \\
2  & 0.3732 & 0.6268 \\
3  & 0.9994 & 0.0006 \\
4  & 0.0611 & 0.9389 \\
5  & 0.7770 & 0.2230 \\
6  & 0.9716 & 0.0284 \\
7  & 0.4111 & 0.5889 \\
8  & 0.1057 & 0.8943 \\
9  & 0.1019 & 0.8981 \\
10 & 0.6094 & 0.3906 \\
11 & 0.9928 & 0.0072 \\
12 & 0.9987 & 0.0013 \\
15 & 0.0006 & 0.9994 \\
\bottomrule
\end{tabular}
\end{table}
Table~\ref{tab:theta_status} reports the estimated trade off weights across AIS navigational status categories obtained from the model fit on the full dataset. These estimates represent the calibrated risk trade offs associated with different navigational status, while uncertainty is assessed separately using bootstrap resampling. The full dataset estimates reported in Table~\ref{tab:theta_status} are broadly consistent with the bootstrap 95\% confidence intervals presented in Table~\ref{tab:status_theta}. In most cases, the point estimates from the main fit lie within the corresponding bootstrap confidence intervals, indicating stability in the inferred trade offs across navigational status.

For certain statuses, such as \textit{aground} (Status 6) and \textit{not defined} (Status 15), the bootstrap intervals are extremely narrow and concentrated near the boundaries of the parameter space, indicating strongly identified behavior dominated by environmental (ice related) risk. Similarly, statuses such as \textit{under way using engine} (Status 0) and \textit{constrained by her draught} (Status 4) exhibit relatively tighter intervals, suggesting more stable and well identified trade offs.

In contrast, several statuses including \textit{at anchor} (Status 1), \textit{restricted maneuverability} (Status 3), \textit{moored} (Status 5), and \textit{engaged in fishing} (Status 7) display wide confidence intervals that span much of the feasible parameter space. Although the full dataset estimates for these categories suggest higher weighting on whale related risk, the associated uncertainty indicates that these trade offs are less precisely identified.

Intermediate statuses, such as \textit{reserved for future amendment} (Status 10) and \textit{under way sailing} (Status 8), exhibit both moderate point estimates and wide confidence intervals, reflecting substantial variability in inferred behavior.

Overall, the comparison indicates that while the full dataset model provides consistent point estimates across navigational status, the degree of uncertainty varies considerably.

\subsection{Model Implied Optimal Speed and Validation}

Using the estimated parameters from the full dataset fit, we computed the model implied optimal speed for each observation by minimizing the latent risk function over the admissible speed grid, \(v \in [0, v_{\max}]\). This was carried out separately for the vessel group and navigational status specifications using the corresponding fitted trade off weights. The resulting optimal speeds provided a validation check by comparing observed segment speed with the speed implied by the estimated decision model.

\subsubsection{Optimal Speed under the Vessel Group Specification}

\begin{figure}[h]
\centering
\includegraphics[width=.95\textwidth]{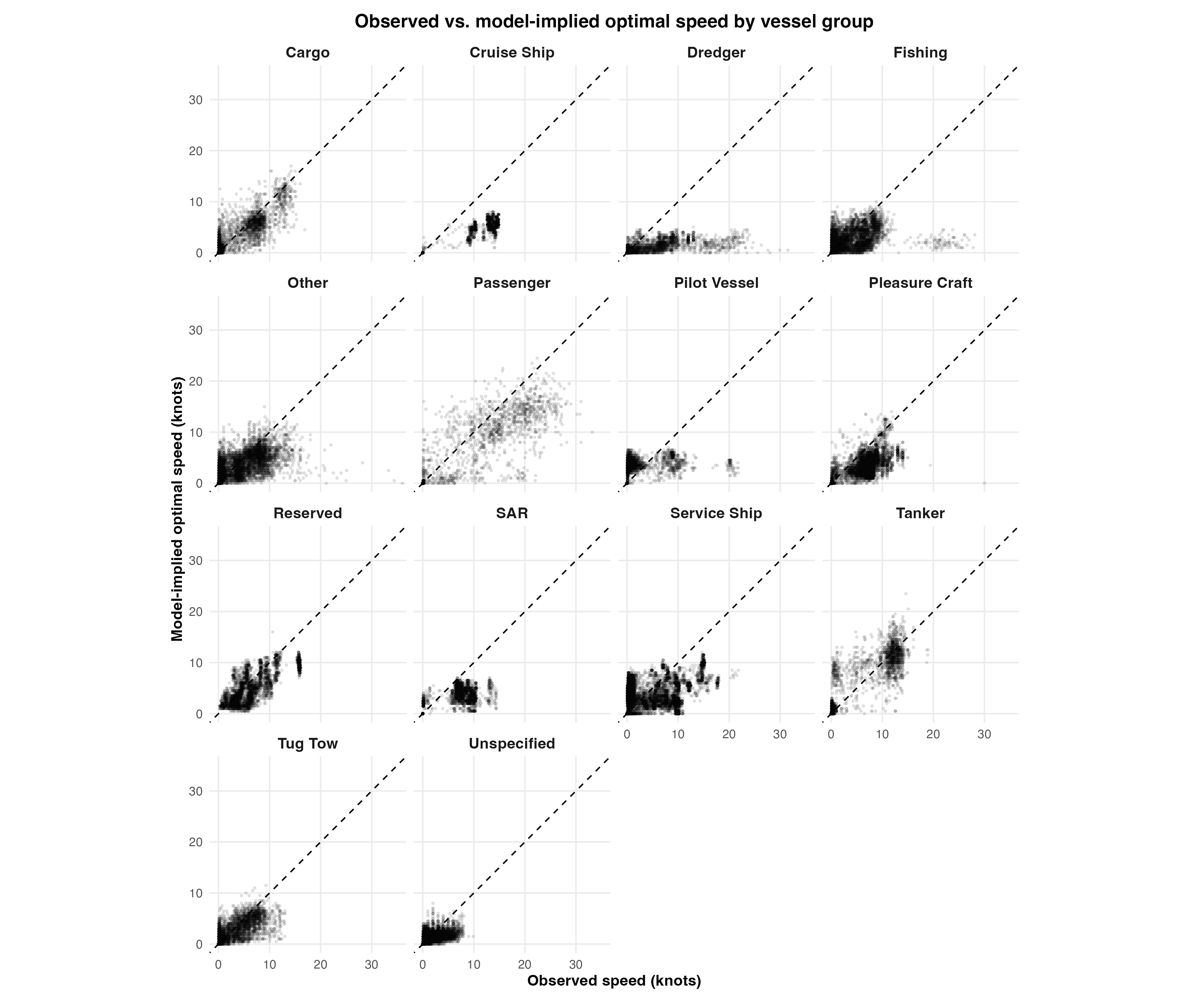}
\caption{Observed versus model implied optimal speed across vessel groups. Each panel compares observed segment speed with the optimal speed obtained by minimizing the fitted latent risk function under the vessel group specification. The dashed The dashed $45^\circ$ line indicates perfect agreement.}
\label{fig:vopt_vg}
\end{figure}

Under the vessel group specification, the correlation between observed segment speed and model implied optimal speed was 0.81($R^2=.66$, indicating strong overall agreement between the inferred decision model and observed vessel speed. Figure~\ref{fig:vopt_vg} compares observed and model implied optimal speeds across vessel groups, while Table~\ref{tab:vopt_vg_summary} summarizes the corresponding group level correlations and mean speeds.

Agreement varies across vessel categories. High correlations are observed for passenger (\(r=0.92\)), tanker (\(r=0.92\)), cargo (\(r=0.89\)), and cruise ship (\(r=0.88\)) vessels, suggesting that the estimated trade off structure closely matches observed speed in these groups. Tug tow vessels, which account for the majority of observations, also exhibit strong agreement (\(r=0.80\)). Moderate correlations are found for other vessel groups, including reserved (\(r=0.75\)), other (\(r=0.72\)), pleasure craft (\(r=0.68\)), unspecified (\(r=0.67\)), dredger (\(r=0.65\)), and fishing vessels (\(r=0.59\)). Lower correlations for pilot vessels (\(r=0.31\)), service ships (\(r=0.28\)), and SAR vessels (\(r=0.47\)) suggest that speed choices in these categories may be influenced by additional operational factors not fully captured by the current risk specification.

\begin{table}[htbp]
\centering
\caption{Observed and model implied optimal speed under the vessel group specification.}
\label{tab:vopt_vg_summary}
\begin{tabular}{lrrr}
\toprule
\textbf{Vessel Group} & \textbf{Mean Observed} & \textbf{Mean Optimal} & \textbf{Correlation} \\
\midrule
Cargo          & 1.30 & 1.39 & 0.89 \\
Cruise Ship    & 10.61 & 4.46 & 0.88 \\
Dredger        & 3.01 & 0.77 & 0.65 \\
Fishing        & 2.07 & 1.43 & 0.59 \\
Other          & 2.74 & 2.22 & 0.72 \\
Passenger      & 1.90 & 1.37 & 0.92 \\
Pilot Vessel   & 2.25 & 2.97 & 0.31 \\
Pleasure Craft & 5.93 & 2.72 & 0.68 \\
Reserved       & 6.50 & 4.82 & 0.75 \\
SAR            & 6.65 & 3.52 & 0.47 \\
Service Ship   & 4.38 & 4.35 & 0.28 \\
Tanker         & 1.67 & 1.84 & 0.92 \\
Tug Tow        & 1.24 & 1.09 & 0.80 \\
Unspecified    & 1.03 & 0.76 & 0.67 \\
\bottomrule
\end{tabular}
\end{table}

\subsubsection{Optimal Speed under the Navigational Status Specification}

\begin{figure}[htbp]
\centering
\includegraphics[width=.95\textwidth]{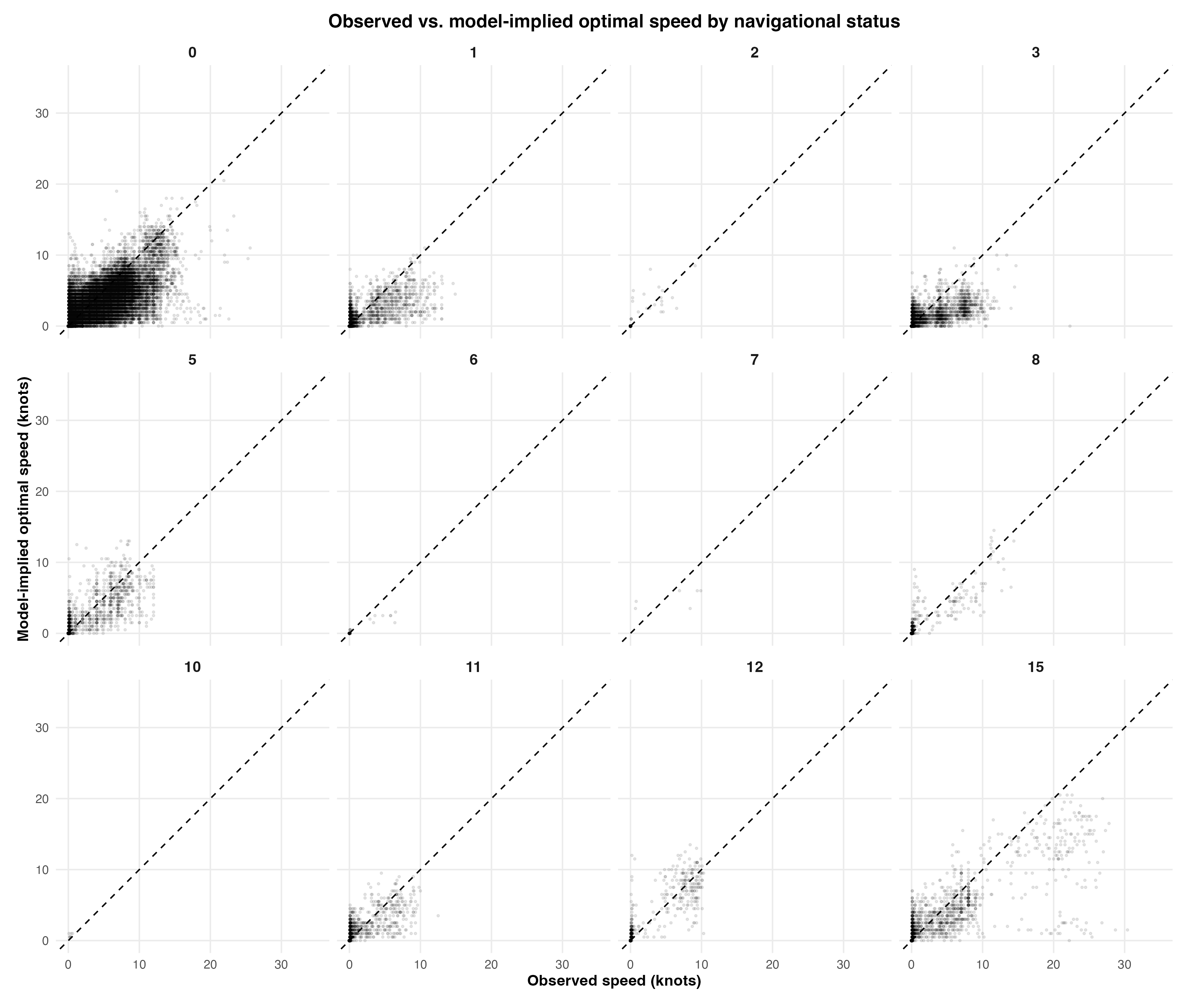}
\caption{Observed versus model implied optimal speed across AIS navigational status codes. Each panel corresponds to a distinct operational state under the navigational status specification. The dashed 45 degree line indicates perfect agreement between observed and model implied speeds. Strong alignment is observed for several dominant status, while greater dispersion in other categories reflects increased behavioral heterogeneity and weaker model fit in specialized operational contexts.}
\label{fig:vopt_st}
\end{figure}

Under the navigational status specification, the correlation between observed segment speed and model implied optimal speed was also 0.81, indicating similarly strong overall agreement between the inferred decision model and observed vessel behavior. Figure~\ref{fig:vopt_st} compares observed and model implied optimal speeds across navigational status, while Table~\ref{tab:vopt_st_summary} summarizes the status level correlations and mean speeds.

At the level of individual vessel status, model fit varies across statuses. High correlations are observed for dominant statuses such as code 0 (\textit{under way using engine}, \(r=0.82\)), code 5 (\textit{moored}, \(r=0.80\)), code 8 (\textit{under way sailing}, \(r=0.80\)), code 12 (\textit{reserved (12)}, \(r=0.86\)), and code 15 (\textit{undefined}, \(r=0.86\)). Moderate agreement is observed for code 1 (\textit{at anchor}, \(r=0.71\)) and code 3 (\textit{restricted maneuverability}, \(r=0.63\)). Lower correlations for more specialized or sparsely represented categories, such as code 7 (\textit{engaged in fishing}, \(r=0.39\)) and code 10 (\textit{reserved (10)}, \(r=0.07\)), indicate that behavior in these contexts may be driven by additional factors not fully captured by the current formulation. Categories with extremely small sample sizes should be interpreted cautiously.

\begin{table}[htbp]
\centering
\caption{Observed and model implied optimal speed under the navigational status specification.}
\label{tab:vopt_st_summary}
\begin{tabular}{r r r r}
\toprule
\textbf{Status} & \textbf{Mean Observed} & \textbf{Mean Optimal} & \textbf{Correlation} \\
\midrule
0  & 1.52 & 1.32 & 0.82 \\
1  & 0.70 & 0.70 & 0.71 \\
2  & 0.47 & 0.42 & 0.85 \\
3  & 1.73 & 1.06 & 0.63 \\
4  & 5.70 & 2.00 &     \\
5  & 1.16 & 1.04 & 0.80 \\
6  & 0.60 & 0.36 & 0.86 \\
7  & 5.52 & 5.34 & 0.39 \\
8  & 1.50 & 1.51 & 0.80 \\
9  & 5.10 & 1.50 &     \\
10 & 0.28 & 0.88 & 0.07 \\
11 & 0.65 & 0.71 & 0.77 \\
12 & 1.94 & 2.19 & 0.86 \\
15 & 1.49 & 1.13 & 0.86 \\
\bottomrule
\end{tabular}
\begin{flushleft}
{\footnotesize
AIS navigation status codes follow the standard AIS specification: Status~0 = under way using engine; Status~1 = at anchor; Status~2 = not under command; Status~3 = restricted maneuverability; Status~4 = constrained by her draught; Status~5 = moored; Status~6 = aground; Status~7 = engaged in fishing; Status~8 = under way sailing; Status~9 = reserved for future amendment; Status~10 = power driven vessel towing astern; Status~11 = power driven vessel towing alongside; Status~12 = reserved for future use; Status~13 = reserved for future use; Status~14 = AIS SART (search and rescue transmitter); Status~15 = not defined. Rare navigation status categories ($<0.05\%$) are retained for completeness.
}
\end{flushleft}
\end{table}

\subsection{Bootstrap Based Uncertainty Quantification}

Uncertainty estimates are obtained from 300 bootstrap replications using a stratified bootstrap procedure \citep{Efron1979, Efron1986}, each based on a stratified sample of approximately one million observations. Stratification is performed by vessel group and navigational status to preserve the underlying composition of the data. Within each bootstrap replicate, pseudo trajectories are sampled with replacement and all associated observations are retained, preserving within trajectory dependence while maintaining the approximate distribution of observations across groups. The resulting bootstrap distribution is used to quantify uncertainty in the inferred trade offs for both the vessel group and status model specification. To improve computational efficiency and numerical stability, each bootstrap optimization is initialized using a warm start based on parameter estimates from the full dataset model fit.

The bootstrap intervals reveal substantial variation in the precision with which trade offs are identified across vessel groups. 
For some groups, such as fishing, passenger, and unspecified vessels, the estimated weights are tightly concentrated near the 
boundaries of the simplex, with extremely narrow confidence intervals. This indicates strong and stable identification of behavior 
dominated by environmental (ice related) risk.

Similarly, vessel groups such as tanker and pleasure craft exhibit consistently high values of $\hat{\theta}_W$ with relatively 
tight confidence intervals, indicating robust and well identified sensitivity to whale related risk.

In contrast, vessel groups such as cruise ships, pilot vessels, SAR, and service vessels exhibit wide confidence intervals 
that span much of the feasible parameter space. This suggests either substantial heterogeneity in behavior within these categories 
or limited information in the data to precisely recover trade offs.

Intermediate groups, including cargo, other, and dredger vessels, display moderate uncertainty but maintain a consistent directional 
pattern toward ice dominated behavior. Tug tow vessels, which account for the majority of observations, exhibit a well identified 
interior solution with relatively narrow confidence intervals centered near equal weighting, indicating a stable balance between 
whale related and ice related risk.

Overall, these results highlight that both the magnitude and the precision of inferred trade offs vary across vessel groups. 
While some vessel types exhibit strongly identified and stable behavioral patterns, others reflect either heterogeneous decision making 
or weaker identifiability, underscoring the importance of uncertainty quantification in interpreting inverse optimization results.

\begin{table}[htbp]
\centering
\caption{Estimated trade off weights across vessel groups with bootstrap confidence intervals.}
\label{tab:vessel_theta}
\begin{tabular}{lcc}
\toprule
\textbf{Vessel Group} & \textbf{$\hat{\theta}_W$ (CI)} & \textbf{$\hat{\theta}_I$ (CI)} \\
\midrule
Cargo          & 0.22 [0.00, 0.43] & 0.78 [0.57, 1.00] \\
Cruise Ship    & 0.38 [0.00, 1.00] & 0.62 [0.00, 1.00] \\
Dredger        & 0.04 [0.00, 0.53] & 0.96 [0.47, 1.00] \\
Fishing        & 0.00 [0.00, 0.00] & 1.00 [1.00, 1.00] \\
Other          & 0.16 [0.00, 0.57] & 0.84 [0.43, 1.00] \\
Passenger      & 0.00 [0.00, 0.00] & 1.00 [1.00, 1.00] \\
Pilot Vessel   & 0.18 [0.00, 0.99] & 0.82 [0.01, 1.00] \\
Pleasure Craft & 0.98 [0.76, 1.00] & 0.02 [0.00, 0.24] \\
Reserved       & 0.90 [0.06, 1.00] & 0.10 [0.00, 0.94] \\
SAR            & 0.46 [0.00, 1.00] & 0.54 [0.00, 1.00] \\
Service Ship   & 0.49 [0.00, 1.00] & 0.51 [0.00, 1.00] \\
Tanker         & 0.95 [0.70, 1.00] & 0.05 [0.00, 0.30] \\
Tug Tow        & 0.50 [0.28, 0.72] & 0.50 [0.28, 0.72] \\
Unspecified    & 0.01 [0.00, 0.06] & 0.99 [0.94, 1.00] \\
\bottomrule
\end{tabular}
\end{table}

The bootstrap intervals reveal substantial variation in the precision of estimated trade offs across navigational status categories. 
Certain statuses exhibit tightly concentrated estimates near the boundaries of the simplex, indicating strongly identified and stable behavior. 
In particular, status 15 (\textit{undefined}) is consistently dominated by environmental (ice related) risk, while status 11 exhibits 
near complete weighting on whale related risk. Both categories display extremely narrow confidence intervals, indicating highly stable 
and well identified trade offs.

Status 0 (\textit{under way using engine}), which accounts for the majority of observations, exhibits a relatively tight confidence interval 
centered away from the boundaries. This suggests a well identified interior solution, with a clear tendency toward ice dominated behavior 
while still maintaining a meaningful trade off between whale related and environmental risks.

In contrast, many statuses including \textit{at anchor} (1), \textit{not under command} (2), \textit{moored} (5), \textit{aground} (6), 
\textit{engaged in fishing} (7), and \textit{under way sailing} (8) display wide confidence intervals that span most of the feasible 
parameter space. This indicates that the data provide limited information to precisely identify trade offs in these operational contexts, 
likely reflecting heterogeneous or context dependent behavior.

Several intermediate statuses, such as \textit{constrained by draught} (4), \textit{reserved (9)}, and \textit{reserved (10)}, show moderate 
uncertainty but exhibit a consistent tendency toward environmental (ice related) risk dominance. Conversely, statuses such as 
\textit{restricted maneuverability} (3) and \textit{reserved (12)} display higher mean values of $\hat{\theta}_W$, suggesting greater 
sensitivity to whale related risk, although their confidence intervals remain relatively wide.

Overall, these results indicate that while certain navigational status exhibit statistically significant risk and narrow uncertainty confidence intervals, many 
statuses are characterized by substantial uncertainty. This reflects either variability in operational conditions within these categories 
or weaker identification of risk trade offs in the available data. Compared to the vessel group specification, the navigational status model 
exhibits greater uncertainty overall, suggesting that vessel type provides a more stable basis for identifying behavioral trade offs than 
transient operational status.
\begin{table}[h]
\centering
\caption{Estimated trade off weights across AIS navigational status categories with bootstrap confidence intervals.}
\label{tab:status_theta}
\begin{tabular}{lcc}
\toprule
\textbf{Status} & \textbf{$\hat{\theta}_W$ (CI)} & \textbf{$\hat{\theta}_I$ (CI)} \\
\midrule
0  & 0.22 [0.05, 0.38] & 0.78 [0.62, 0.95] \\
1  & 0.45 [0.00, 1.00] & 0.55 [0.00, 1.00] \\
2  & 0.26 [0.00, 0.99] & 0.74 [0.01, 1.00] \\
3  & 0.88 [0.00, 1.00] & 0.12 [0.00, 1.00] \\
4  & 0.12 [0.00, 0.77] & 0.88 [0.23, 1.00] \\
5  & 0.65 [0.00, 1.00] & 0.35 [0.00, 1.00] \\
6  & 0.66 [0.00, 1.00] & 0.34 [0.00, 1.00] \\
7  & 0.72 [0.00, 1.00] & 0.28 [0.00, 1.00] \\
8  & 0.24 [0.00, 1.00] & 0.76 [0.00, 1.00] \\
9  & 0.14 [0.00, 0.83] & 0.86 [0.17, 1.00] \\
10 & 0.41 [0.00, 0.97] & 0.59 [0.03, 1.00] \\
11 & 1.00 [0.99, 1.00] & 0.00 [0.00, 0.01] \\
12 & 0.94 [0.40, 1.00] & 0.06 [0.00, 0.60] \\
15 & 0.00 [0.00, 0.00] & 1.00 [1.00, 1.00] \\
\bottomrule
\end{tabular}
\end{table}



\section{Sensitivity Analysis}

We evaluate alternative plausible scenarios that alter the relative importance of whale related and ice related risk in the inferred objective function. In all cases, weights are renormalized to preserve the simplex constraint.

\paragraph{Whale risk perturbations.}
We conduct a sensitivity analysis in which whale related risk is doubled, with weights renormalized to preserve the simplex constraint. Across vessel groups, the resulting changes in model implied optimal speed are very small, and median effects are zero in all categories. Mean changes are generally on the order of $10^{ 3}$ to $10^{ 2}$ knots. The largest increases are observed for service ships (0.09 knots), cargo vessels (0.03 knots), and cruise ships (0.03 knots), while only very small decreases occur for unspecified vessels ($ 0.002$ knots), pilot vessels ($ 0.0005$ knots), and dredgers ($ 0.0002$ knots). Overall, doubling the whale risk weight produces only negligible changes in optimal speed and no consistent tendency toward slower travel.


\paragraph{Ice risk perturbations.}
In contrast, increasing the weight on ice related risk by a factor of two produces more systematic reductions in model implied optimal speed across vessel groups. Median changes remain zero across all groups, indicating that most observations are unaffected, but mean changes are predominantly negative. The largest decreases are observed for service ships ($ 0.06$ knots), reserved vessels ($ 0.03$ knots), tug tow vessels ($ 0.02$ knots), and cargo vessels ($ 0.02$ knots), with smaller reductions for cruise ships, other vessels, tankers, SAR, pleasure craft, and pilot vessels. Only negligible increases are observed for dredgers (0.0003 knots) and unspecified vessels (0.0009 knots). Compared with the whale risk perturbations, the ice risk perturbation yields a much more consistent directional response, indicating that ice related constraints exert a stronger and more systematic influence on vessel speed decisions.

\paragraph{Status level comparison.}
A similar asymmetry is observed across navigational status categories. Increasing the whale risk weight leads to only modest changes in optimal speed, typically below 0.05 knots, and in several cases results in small increases rather than reductions. For example, vessels under way using engine (code 0) and reserved (code 10) exhibit increases of approximately 0.03  0.05 knots under the amplified whale risk scenario.

In contrast, increasing the ice risk weight produces more consistent reductions in optimal speed across statuses. Negative adjustments are observed for most categories, including vessels under way using engine (code 0), at anchor (code 1), engaged in fishing (code 7), under way sailing (code 8), and reserved (code 10), with mean reductions generally on the order of 0.01  0.02 knots. As in the vessel group analysis, median effects are zero across all categories, reflecting the discrete speed grid and localized adjustments.

\paragraph{Implications.}
These results reveal a clear asymmetry in the inferred decision structure. Ice related constraints exert a stronger and more systematic influence on vessel speed than whale related risk. While increasing whale risk weighting alone does not lead to meaningful reductions in speed, increasing ice related constraints produces consistent reductions in optimal vessel speed. This suggests that vessel operator speed choice is primarily governed by operational and environmental constraints associated with ice conditions, whereas whale related risk, in isolation, does not act as a binding constraint. Consequently, policies that rely solely on tying increased whale risk penalties to speed may have limited effectiveness, whereas interventions that directly affect operational constraints are more likely to influence vessel operator decisions.
\section{Discussion}

This study develops a data driven inverse control constrained optimization framework to infer how vessel operators balance competing risks under complex Arctic conditions. By combining AIS data with environmental covariates and spatial whale intensity estimates, our analysis recovers interpretable trade offs governing vessel speed decisions.

Three distinct risk regimes emerge across vessel groups. Fishing, passenger, and unspecified vessels exhibit strongly identified boundary solutions dominated by ice related risk, with point estimates near $\hat{\theta}_I \approx 1$ and extremely narrow bootstrap confidence intervals. Validation and policy sensitivity analyses support this interpretation, showing that variation in whale related risk has minimal influence on optimal speed, whereas ice related risk produces more pronounced adjustments. The spatial patterns help explain this result, but differ across groups. Fishing vessels operate along a persistent corridor in the Chukchi and Beaufort Seas where ice concentration is consistently high (Figure \ref{fig:appendix_group_fishing_other}a), leading to a clear and well supported dominance of ice related constraints. In contrast, passenger and unspecified vessels exhibit much more limited and spatially concentrated activity (Figures \ref{fig:appendix_group_passenger_pilot}a and \ref{fig:appendix_group_tanker_unspecified}b), with sparse observations and minimal overlap with regions of elevated whale intensity. In these cases, the lack of co occurrence between vessel activity and whale presence further limits the identification of whale related trade offs. Across all three groups, the observed operating environments are effectively dominated by ice risk, implying that whale related risk does not constitute a binding constraint within the decision space.

In contrast, pleasure craft and tanker vessels exhibit tightly identified solutions with high weight on whale related risk. Despite differences in scale and operational purpose, both groups operate in regions with greater spatial overlap between vessel activity and whale presence (Figures \ref{fig:appendix_group_pleasure_reserved} and \ref{fig:appendix_group_tanker_unspecified}), particularly along nearshore and transit corridors. This co occurrence introduces a meaningful trade off between navigation and ecological exposure, leading to elevated estimates of whale related risk. However, sensitivity analysis results indicate that increases in ice related risk still produce stronger reductions in optimal speed, suggesting that while whale risk is behaviorally salient, ice conditions continue to impose binding operational constraints.

\begin{figure}[htbp]
\centering
\subfloat[Tug Tow]{
\includegraphics[width=0.95\textwidth]{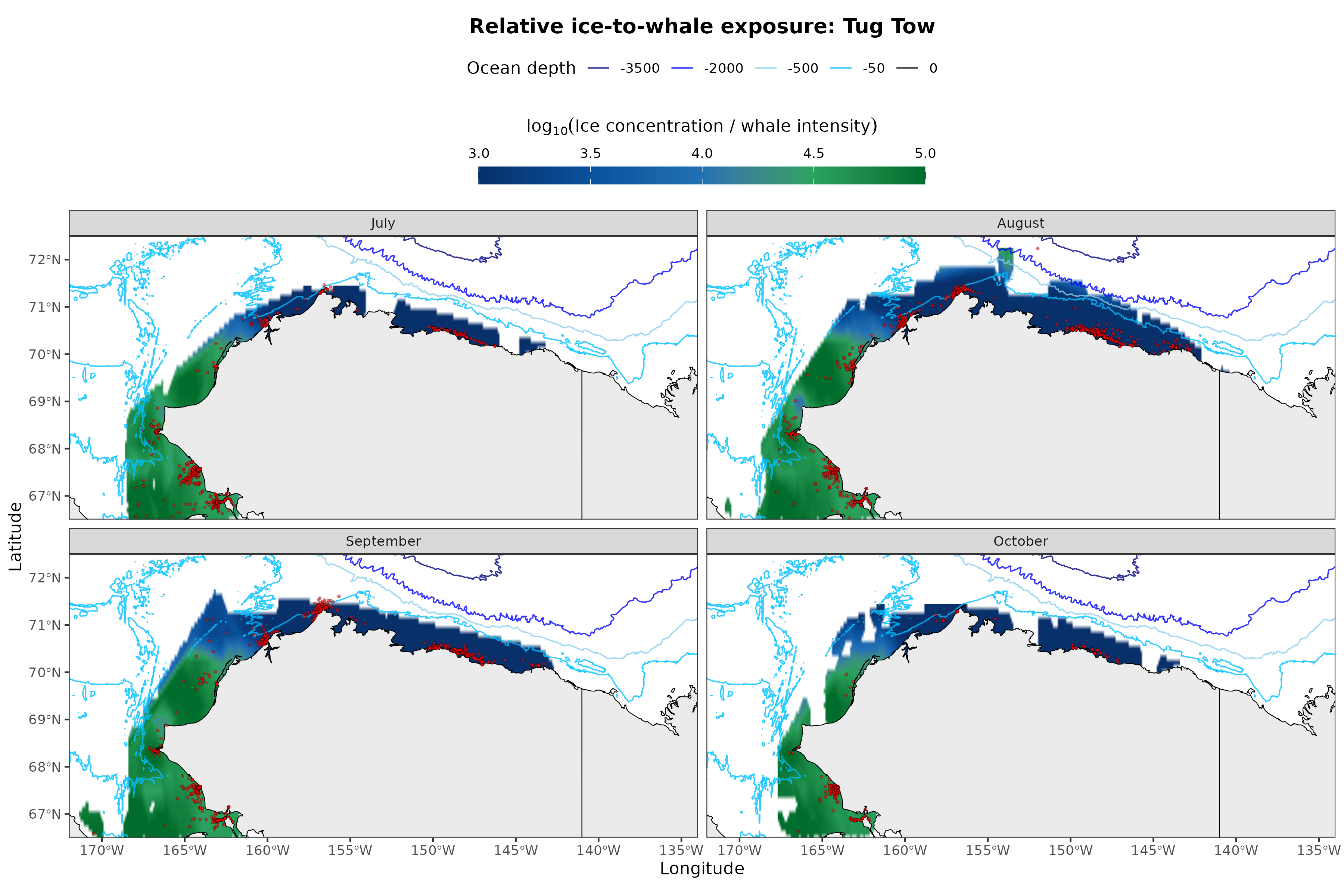}
}

\vspace{0.4cm}

\subfloat[Cargo]{
\includegraphics[width=0.95\textwidth]{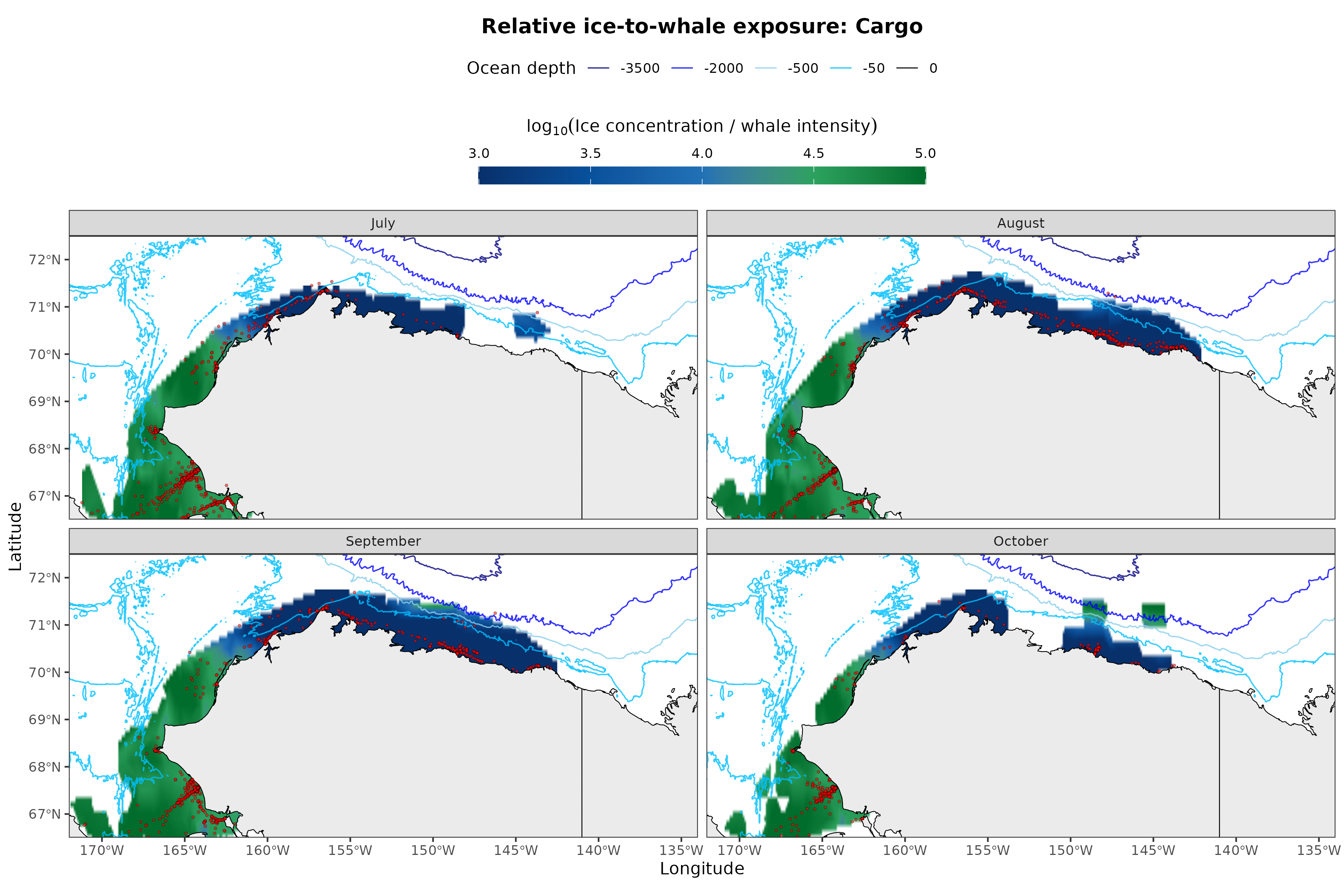}
}

\caption{Monthly spatial distribution of the log ratio of ice concentration to whale intensity.}
\label{fig:appendix_group_tug_tow_cargo}
\end{figure}

Tug tow vessels, which account for the largest share of observations, exhibit a well identified interior solution with a balanced weighting of whale related and ice related risk. This reflects their operation along extended transit corridors where both risks co occur spatially: ice exposure dominates in offshore segments, while whale presence increases closer to coastal regions (Figure \ref{fig:appendix_group_tug_tow_cargo}). As a result, these vessels face a genuine and persistent trade off between navigational constraints and ecological exposure. The relatively narrow confidence intervals and strong agreement between observed and model implied optimal speeds indicate that this trade off is consistently resolved, yielding a stable and well identified behavioral pattern.

These results show that vessel speed decisions are not governed by a single uniform objective. Instead, they reflect heterogeneous responses to environmental and ecological risks across vessel groups, with some groups dominated by ice related constraints, others showing high weighting on whale related risk, and dominant groups such as tug tow operating under a more balanced trade off structure.

A similar analysis was conducted across navigational status categories, although the trade offs are generally less precisely identified compared to vessel groups.

The dominant status, under way using engine (status 0), which accounts for approximately 77\% of observations, exhibits a well identified interior solution with a clear tilt toward ice related risk ($\hat{\theta}_I \approx 0.78$) and relatively tight confidence intervals. The spatial patterns in Figure \ref{fig:appendix_status_0_1}(a) help explain this result: vessel activity spans both the northern ice dominated corridor across the Chukchi and Beaufort Seas (dark blue regions) and more southern coastal areas where whale intensity is relatively higher (green regions). This overlap implies that vessels operating under this status routinely encounter both types of risk, creating a meaningful trade off environment. As a result, speed decisions reflect a balanced but ice leaning response, where ice conditions remain the primary constraint while whale related risk still plays a secondary but non negligible role.

Certain less frequent statuses exhibit strongly identified boundary solutions that reflect highly localized operating domains. Status 15 (\textit{undefined}) is concentrated along the northern transit corridor where ice exposure dominates almost uniformly (dark blue regions) (Figure \ref{fig:appendix_status_15}), explaining its near complete weighting on ice related risk. In contrast, status 11 (\textit{towing alongside}) is spatially confined to southern and near coastal areas (Figure \ref{fig:appendix_status_11}) with relatively higher whale intensity (green regions) and minimal ice exposure. This strong spatial segregation implies that each status operates in environments where one risk overwhelmingly dominates, leaving little effective trade off and leading to boundary solutions with extremely narrow confidence intervals.

In contrast, most other status categories display wide bootstrap confidence intervals spanning much of the feasible parameter space, including common statuses such as \textit{at anchor}, \textit{moored}, and \textit{engaged in fishing}. This indicates substantial heterogeneity or limited information to recover precise trade offs within these categories.

Overall, compared to vessel groups, navigational status provides a less stable basis for identifying behavioral trade offs, likely reflecting the transient and context dependent nature of navigational status.

\subsection{Methodological Contributions and Limitations}

This study contributes to the growing literature on data driven decision inference by demonstrating how inverse optimization can be applied to large scale observational data in a complex environmental setting. By integrating machine learning predictions, spatial ecological models, and optimization based inference, the framework provides both predictive accuracy and operational interpretability.

However, several limitations should be noted. First, the model assumes that vessel speed decisions can be represented as solutions to a static, local optimization problem. While appropriate for AIS based analysis, this abstraction does not capture longer term planning or route level decision making. Second, the risk components are constructed from observable covariates and may omit unobserved factors such as economic incentives, regulatory compliance, or operator specific preferences. Third, uncertainty in environmental inputs, particularly whale intensity estimates, is not fully propagated into the inverse optimization framework.

Future work could extend this framework to incorporate dynamic decision making, additional risk components, and richer representations of uncertainty. Integrating economic cost functions or regulatory constraints may further enhance the interpretability and policy relevance of the inferred trade offs.

\subsection{Implications for Arctic Maritime Systems}

Overall, the results highlight the importance of environmental constraints in shaping vessel behavior in the Arctic. Ice conditions emerge as the dominant factor influencing speed decisions, while ecological risk plays a more limited role under current conditions. At the same time, substantial heterogeneity across vessel groups underscores the need for differentiated policy approaches that account for variation in operational characteristics.

By providing a quantitative framework for recovering latent decision objectives, this study offers a foundation for evaluating how vessel transit may respond to changing environmental conditions, increased traffic, and evolving regulatory frameworks in the Arctic. The approach can also be extended to other transportation systems where decisions are shaped by complex and interacting risk factors.
\section*{Declarations}

\begin{itemize}
\item \textbf{Funding:} This work received no external funding and was conducted as part of the 
author’s doctoral research in Integrative Life Sciences Doctoral Program, Center for Integrative Life Sciences Education, 
Virginia Commonwealth University.
\item \textbf{Conflict of interest/Competing interests:}
The author declares no conflicts of interest.
\item Ethics approval and consent to participate: Not applicable
\item Consent for publication: Not applicable
\item Data availability : Available upon request
\item Materials availability
\item Code availability : Available upon request

\end{itemize}

\noindent

\bibliographystyle{plainnat}
\bibliography{References}%

@article{pallotta2013vessel,
  author  = {Pallotta, Giuliana and Vespe, Michele and Bryan, Karna},
  title   = {Vessel pattern knowledge discovery from AIS data: A framework for anomaly detection and route prediction},
  journal = {Entropy},
  volume  = {15},
  number  = {6},
  pages   = {2218--2245},
  year    = {2013},
  doi     = {10.3390/e15062218}
}

@book{Ross1976,
  author    = {Ross, Donald},
  title     = {Mechanics of Underwater Noise},
  publisher = {Pergamon Press},
  year      = {1976}
}

@article{McKenna2012,
  author  = {McKenna, Megan F. and Ross, Darlene and Wiggins, Sean M. and Hildebrand, John A.},
  title   = {Underwater radiated noise from modern commercial ships},
  journal = {The Journal of the Acoustical Society of America},
  volume  = {131},
  number  = {1},
  pages   = {92--103},
  year    = {2012},
  doi     = {10.1121/1.3664100}
}

@article{bertsimas2015data,
  title={Data-driven estimation in equilibrium using inverse optimization},
  author={Bertsimas, Dimitris and Gupta, Vishal and Paschalidis, Ioannis Ch},
  journal={Mathematical Programming},
  volume={153},
  number={2},
  pages={595--633},
  year={2015},
  publisher={Springer}
}

@article{aswani2018inverse,
  title   = {Inverse Optimization with Noisy Data},
  author  = {Aswani, Anil and Shen, Zuo-Jun Max and Siddiq, Awan},
  journal = {Operations Research},
  volume  = {66},
  number  = {3},
  pages   = {870--892},
  year    = {2018},
  url     = {https://doi.org/10.1287/opre.2017.1705}
}

@inproceedings{ng2000algorithms,
  title     = {Algorithms for Inverse Reinforcement Learning},
  author    = {Ng, Andrew Y. and Russell, Stuart J.},
  booktitle = {Proceedings of the 17th International Conference on Machine Learning (ICML)},
  pages     = {663--670},
  year      = {2000}
}

@inproceedings{abbeel2004apprenticeship,
  title     = {Apprenticeship Learning via Inverse Reinforcement Learning},
  author    = {Abbeel, Pieter and Ng, Andrew Y.},
  booktitle = {Proceedings of the 21st International Conference on Machine Learning (ICML)},
  pages     = {1--8},
  year      = {2004},
  doi       = {10.1145/1015330.1015430}
}

@article{kalman1964when,
  title   = {When Is a Linear Control System Optimal?},
  author  = {Kalman, Rudolf E.},
  journal = {Journal of Basic Engineering},
  volume  = {86},
  number  = {1},
  pages   = {51--60},
  year    = {1964},
  doi     = {10.1115/1.3653115}
}

@article{Montewka2015,
  author  = {Montewka, Jakub and Goerlandt, Floris and Kujala, Pentti and Lensu, Mikko},
  title   = {Towards probabilistic models for the prediction of a ship performance in dynamic ice},
  journal = {Cold Regions Science and Technology},
  volume  = {112},
  pages   = {14--28},
  year    = {2015},
  doi     = {10.1016/j.coldregions.2014.12.009}
}

@article{Psaraftis2013,
  author  = {Psaraftis, Harilaos N. and Kontovas, Christos A.},
  title   = {Speed models for energy-efficient maritime transportation: A taxonomy and survey},
  journal = {Transportation Research Part C},
  volume  = {26},
  pages   = {331--351},
  year    = {2013},
  doi     = {10.1016/j.trc.2012.09.012}
}

@techreport{Berkman2022ArcticTraffic,
  author       = {Berkman, P. A. and Fiske, G. J. and Lorenzini, D. and Young, O. R. and Pletnikoff, K. and Grebmeier, J. M. and Fernandez, L. M. and Divine, L. M. and Causey, D. and Kapsar, K. E. and J{\o}rgensen, L. L.},
  title        = {Satellite Record of Pan-Arctic Maritime Ship Traffic},
  institution  = {NOAA Office of Oceanic and Atmospheric Research},
  series       = {NOAA Technical Report OAR ARC 22-10},
  year         = {2022},
  doi          = {10.25923/mhrv-gr76}
}

@article{ziebart2008maximum,
  title     = {Maximum Entropy Inverse Reinforcement Learning},
  author    = {Ziebart, Brian D. and Maas, Andrew L. and Bagnell, J. Andrew and Dey, Anind K.},
  journal   = {AAAI Conference on Artificial Intelligence},
  volume    = {22},
  number    = {2},
  pages     = {1433--1438},
  year      = {2008}
}

@article{Erbe2012,
  author  = {Erbe, Christine},
  title   = {Underwater noise of small personal watercraft (jet skis)},
  journal = {The Journal of the Acoustical Society of America},
  volume  = {133},
  number  = {4},
  pages   = {EL326--EL330},
  year    = {2013},
  doi     = {10.1121/1.4795220}
}

@article{ahuja2001inverse,
  title={Inverse optimization},
  author={Ahuja, Ravindra K and Orlin, James B},
  journal={Operations Research},
  volume={49},
  number={5},
  pages={771--783},
  year={2001},
  publisher={INFORMS}
}

@article{burton1992inverse,
  title={Inverse shortest paths: constructive algorithms and applications},
  author={Burton, Dan and Toint, Philippe L},
  journal={Mathematical Programming},
  volume={53},
  number={1-3},
  pages={45--58},
  year={1992},
  publisher={Springer}
}

@article{Redfern2013,
  author  = {Redfern, Jessica V. and Moore, T. J. and Fiedler, Paul C. and DeLong, Robert L. and Hurst, Tom P. and Costa, Daniel P.},
  title   = {Predicting cetacean distributions in data-poor marine ecosystems},
  journal = {Diversity and Distributions},
  year    = {2013},
  volume  = {19},
  number  = {3},
  pages   = {364--376},
  doi     = {10.1111/ddi.12537}
}

@article{Silber2014,
  author  = {Silber, Gregory K. and Adams, Jeffrey D. and Bettridge, Sarah and Cole, Thomas V. N. and Couvillion, Samuel P. and Gende, Scott M. and Helm, Robert C. and Jensen, Andrew S. and Karp, William A. and Knowlton, Amy R. and Laist, David W. and Lanfredi, Chris and Leaper, Russell and Lonergan, Matthew and Mattila, David K. and Panigada, Simone and Robbins, Jooke and Slay, Christopher K. and Smith, Timothy D. and Tregenza, Nicola J. C.},
  title   = {The role of the International Maritime Organization in reducing vessel threat to whales: Process, options, action and effectiveness},
  journal = {Marine Policy},
  year    = {2014},
  volume  = {36},
  pages   = {1221--1233},
  doi     = {10.1016/j.marpol.2012.03.008}
}

@article{zhang1996inverse,
  title={A network flow method for solving some inverse combinatorial optimization problems},
  author={Zhang, Jianzhong and Ma, Zhongfan},
  journal={Optimization},
  volume={37},
  number={1},
  pages={59--72},
  year={1996},
  publisher={Taylor \& Francis}
}

@article{Efron1979,
  author  = {Efron, Bradley},
  title   = {Bootstrap Methods: Another Look at the Jackknife},
  journal = {The Annals of Statistics},
  year    = {1979},
  volume  = {7},
  number  = {1},
  pages   = {1--26}
}

@article{dawson2018temporal,
  title={Temporal and spatial patterns of ship traffic in the Canadian Arctic from 1990 to 2015},
  author={Dawson, Jackie and Pizzolato, Larissa and Howell, Stephen EL and Copland, Luke and Johnston, Margaret E},
  journal={Arctic},
  volume={71},
  number={1},
  pages={15--26},
  year={2018},
  publisher={JSTOR}
}

@article{Efron1986,
  author  = {Efron, Bradley and Tibshirani, Robert},
  title   = {Bootstrap Methods for Standard Errors, Confidence Intervals, and Other Measures of Statistical Accuracy},
  journal = {Statistical Science},
  year    = {1986},
  volume  = {1},
  number  = {1},
  pages   = {54--75}
}

@article{Powell2009,
  author  = {Powell, Michael J. D.},
  title   = {The BOBYQA Algorithm for Bound Constrained Optimization Without Derivatives},
  journal = {Cambridge NA Report NA2009/06},
  year    = {2009},
  institution = {Department of Applied Mathematics and Theoretical Physics, University of Cambridge}
}

@article{PANT2026125763,
title = {A gradient boosted mixed-model machine learning framework for vessel speed in the {U.S.} {A}rctic},
journal = {Ocean Engineering},
volume = {359},
pages = {125763},
year = {2026},
issn = {0029-8018},
doi = {10.1016/j.oceaneng.2026.125763},
url = {https://www.sciencedirect.com/science/article/pii/S0029801826015970},
author = {Mauli Pant and Linda Fernandez and Indranil Sahoo}
}

@article{pant2025lgcp,
  title={Spatio-temporal Shared-Field Modeling of Beluga and Bowhead Whale Sightings Using a Joint Marked Log-Gaussian Cox Process},
  author={Pant, Mauli and Fernandez, Linda and Sahoo, Indranil},
  journal={arXiv preprint arXiv:2512.06450},
  year={2025}
}

@article{xing2023fishing,
  title   = {The Study of Fishing Vessel Behavior Identification Based on AIS Data: A Case Study of the East China Sea},
  author  = {Xing, Bowen and Zhang, Liang and Liu, Zhenchong and Sheng, Hengjiang and Bi, Fujia and Xu, Jingxiang},
  journal = {Journal of Marine Science and Engineering},
  volume  = {11},
  number  = {5},
  pages   = {1093},
  year    = {2023},
  doi     = {10.3390/jmse11051093}
}

@techreport{HSVA2014FuelConsumption,
  author      = {{HSVA Arctic Technology}},
  title       = {Calculation of fuel consumption per mile for various ship types and ice conditions in past, present and in future},
  number      = {D2.42},
  institution = {ACCESS Arctic Climate Change, Economy and Society, Seventh Framework Programme (Project \#265863)},
  year        = {2014},
  note        = {Figure~8: Speed of ship depending on ice concentration}
}

\clearpage

\section*{Appendix}

\addcontentsline{toc}{section}{Appendix}

\setcounter{section}{0}

\renewcommand{\thesection}{A}

\section{Vessel Group Ratio Plots}\label{secA1}

To provide spatial context for the inferred behavioral trade offs, we construct monthly maps of the relative exposure to ice and whale related risk across vessel groups. For each observation, we compute a localized measure of environmental exposure by aggregating sea ice concentration and whale intensity onto a spatial grid. We then define the ratio of ice concentration to whale intensity and visualize its logarithm, $\log_{10}\!\left(\frac{\text{ice concentration}}{\text{whale intensity}}\right)$, which provides a normalized measure of the dominant environmental constraint. Higher values indicate regions where ice related exposure dominates, while lower values correspond to areas with relatively greater whale presence.

To reduce noise and ensure interpretability, the ratio is averaged within spatial grid cells and months, trimmed to remove extreme values, and interpolated over the observed data of vessel locations. Interpolation is restricted to regions with sufficient data density and further constrained using a convex hull of observed vessel locations to avoid extrapolation into unsupported areas. The resulting maps therefore reflect the effective operational environment experienced by each vessel group. Overlaid points indicate observed vessel locations, allowing direct comparison between spatial exposure patterns and vessel activity.

\begin{figure}[htbp]
\centering
\subfloat[Cruise Ship]{
\includegraphics[width=0.95\textwidth]{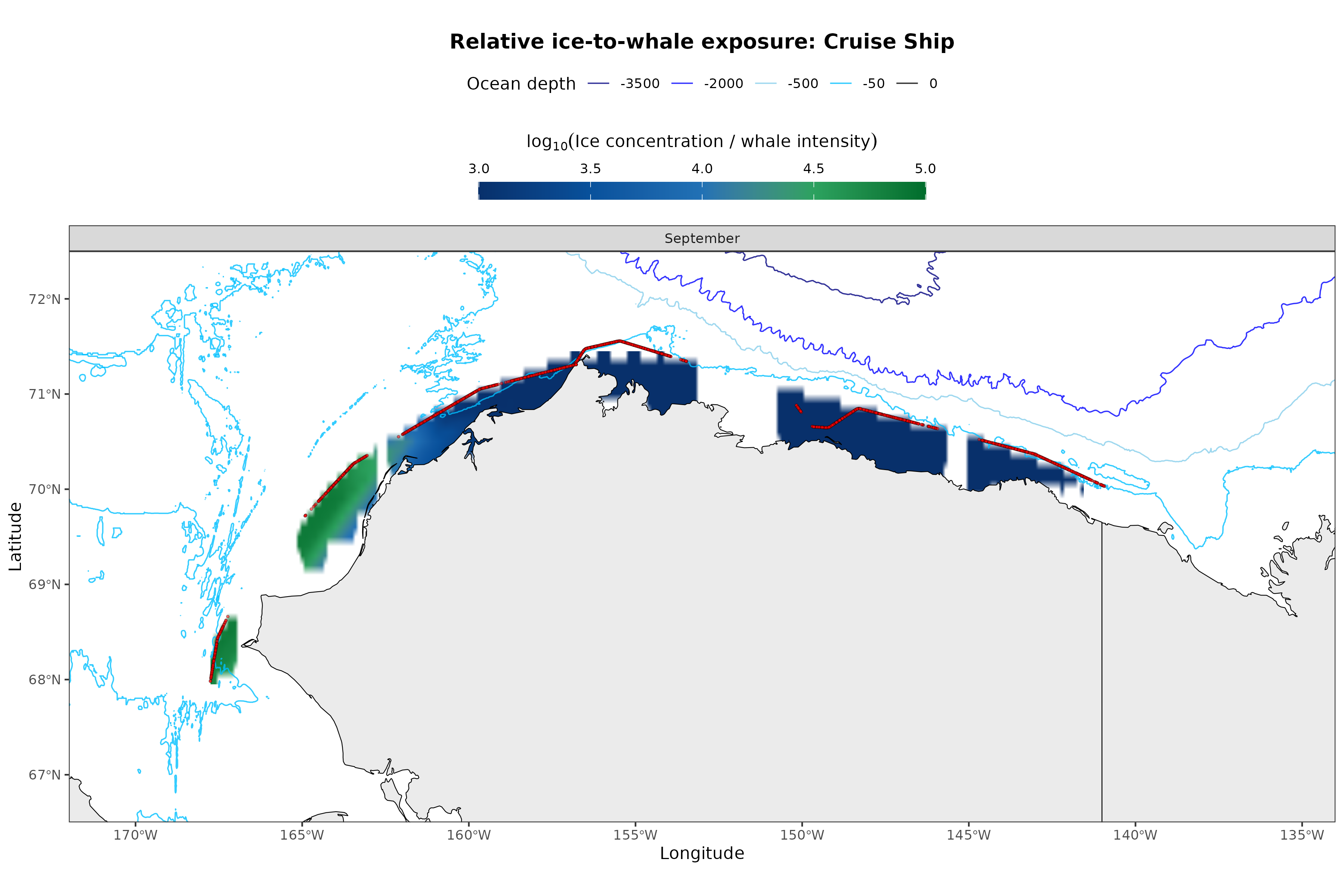}
}

\vspace{0.4cm}

\subfloat[Dredger]{
\includegraphics[width=0.95\textwidth]{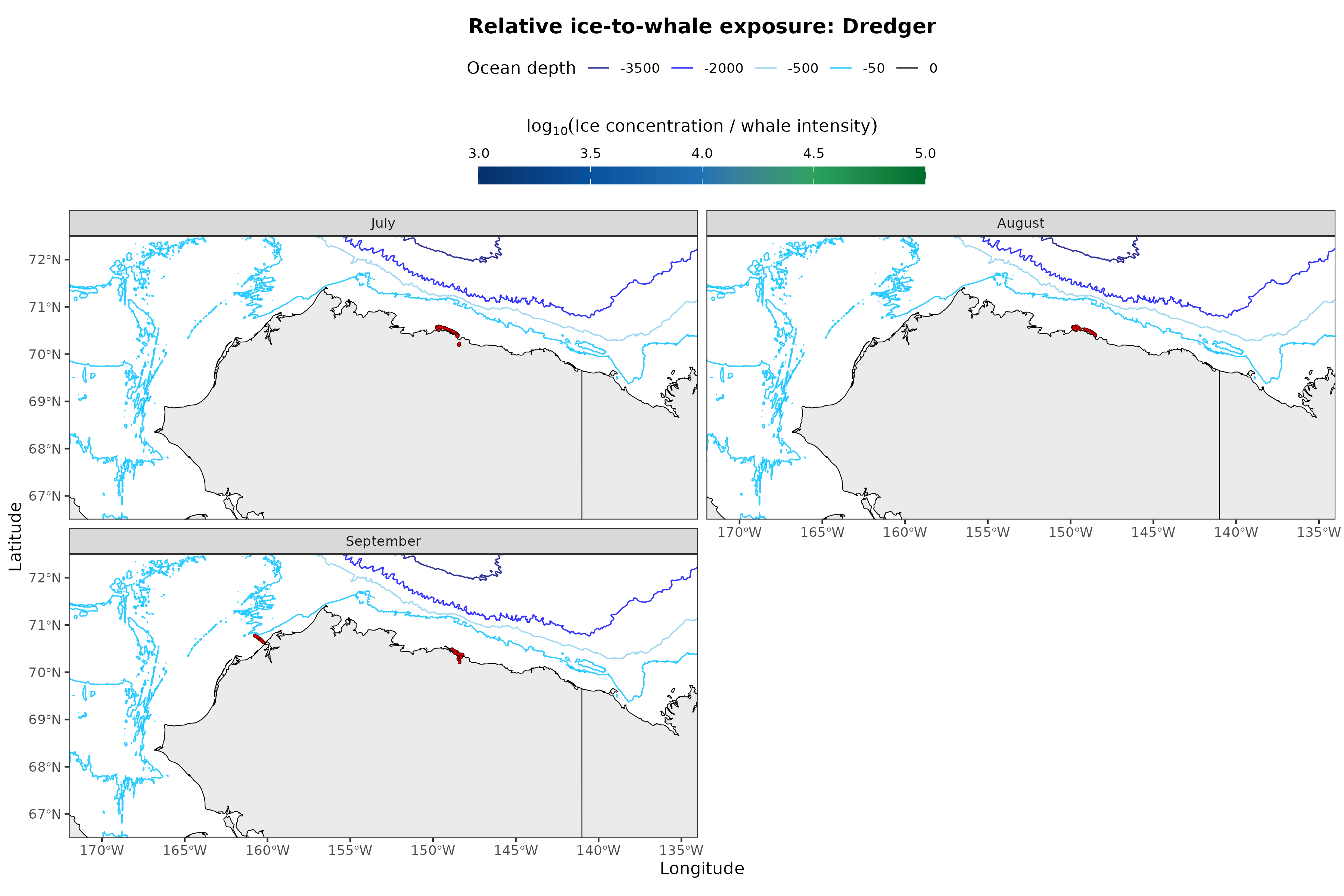}
}

\caption{Monthly spatial distribution of the log ratio of ice concentration to whale intensity.}
\label{fig:appendix_group_cruise_dredger}
\end{figure}

\begin{figure}[htbp]
\centering
\subfloat[Fishing]{
\includegraphics[width=0.95\textwidth]{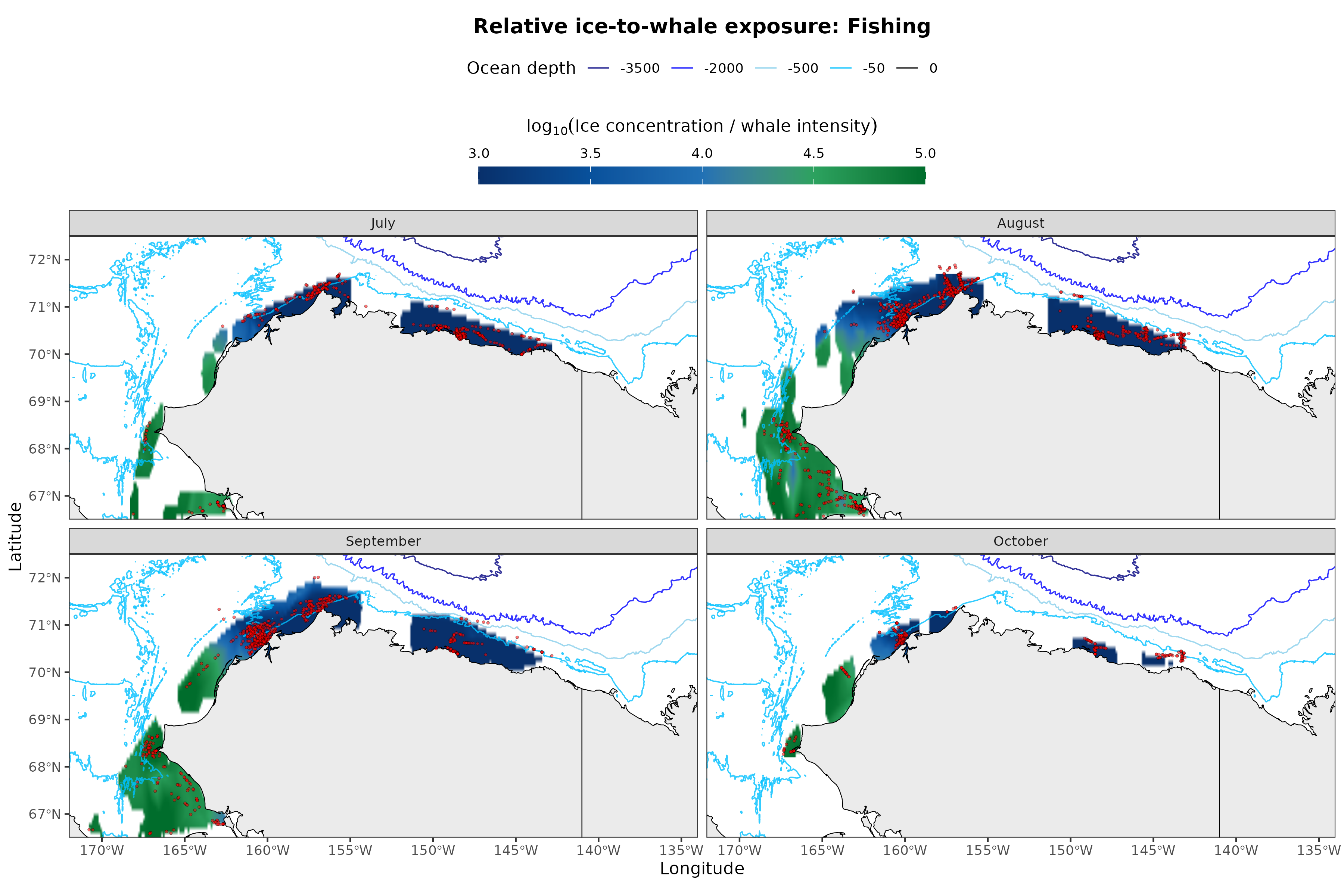}
}

\vspace{0.4cm}

\subfloat[Other]{
\includegraphics[width=0.95\textwidth]{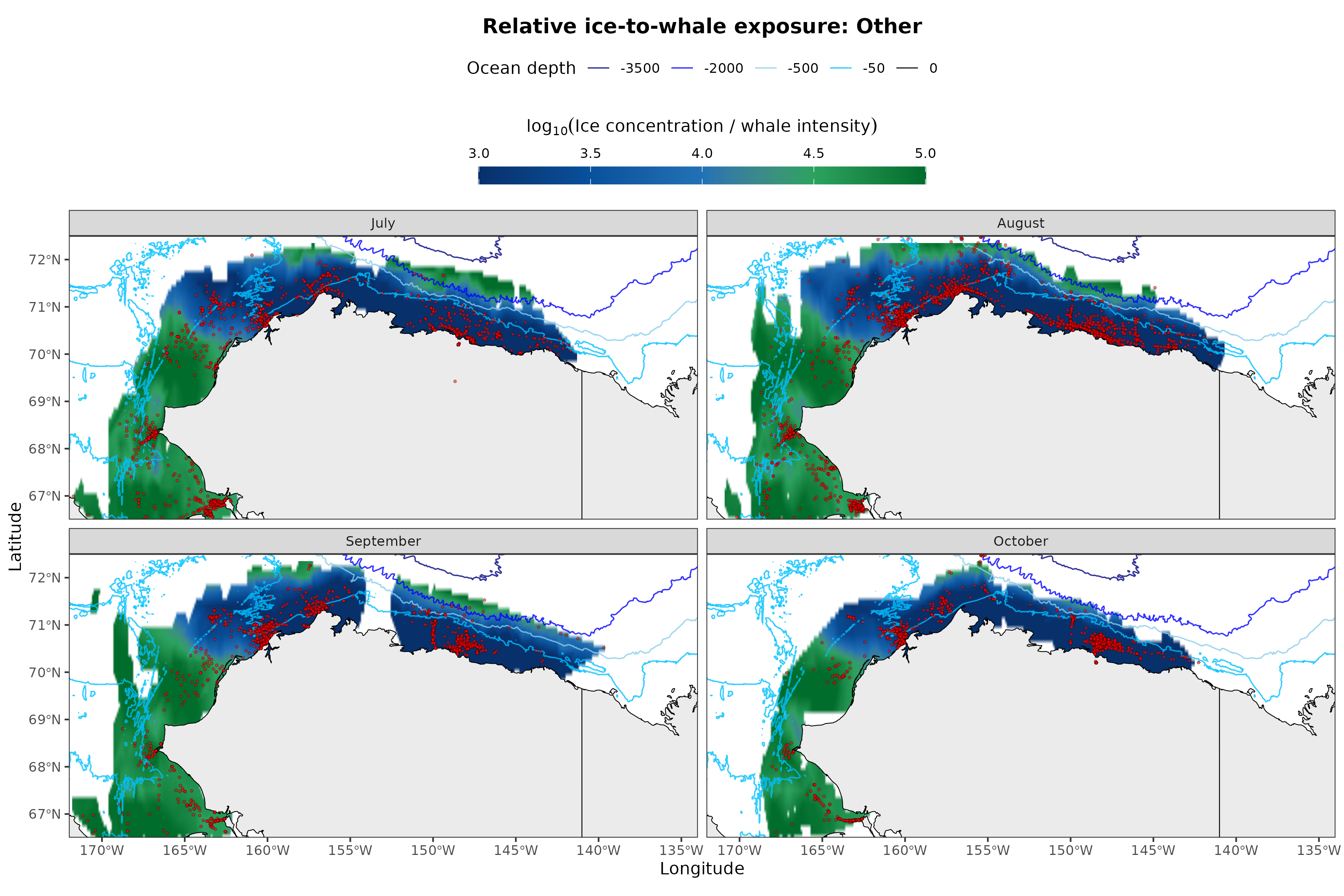}
}

\caption{Monthly spatial distribution of the log ratio of ice concentration to whale intensity.}
\label{fig:appendix_group_fishing_other}
\end{figure}

\begin{figure}[htbp]
\centering
\subfloat[Passenger]{
\includegraphics[width=0.95\textwidth]{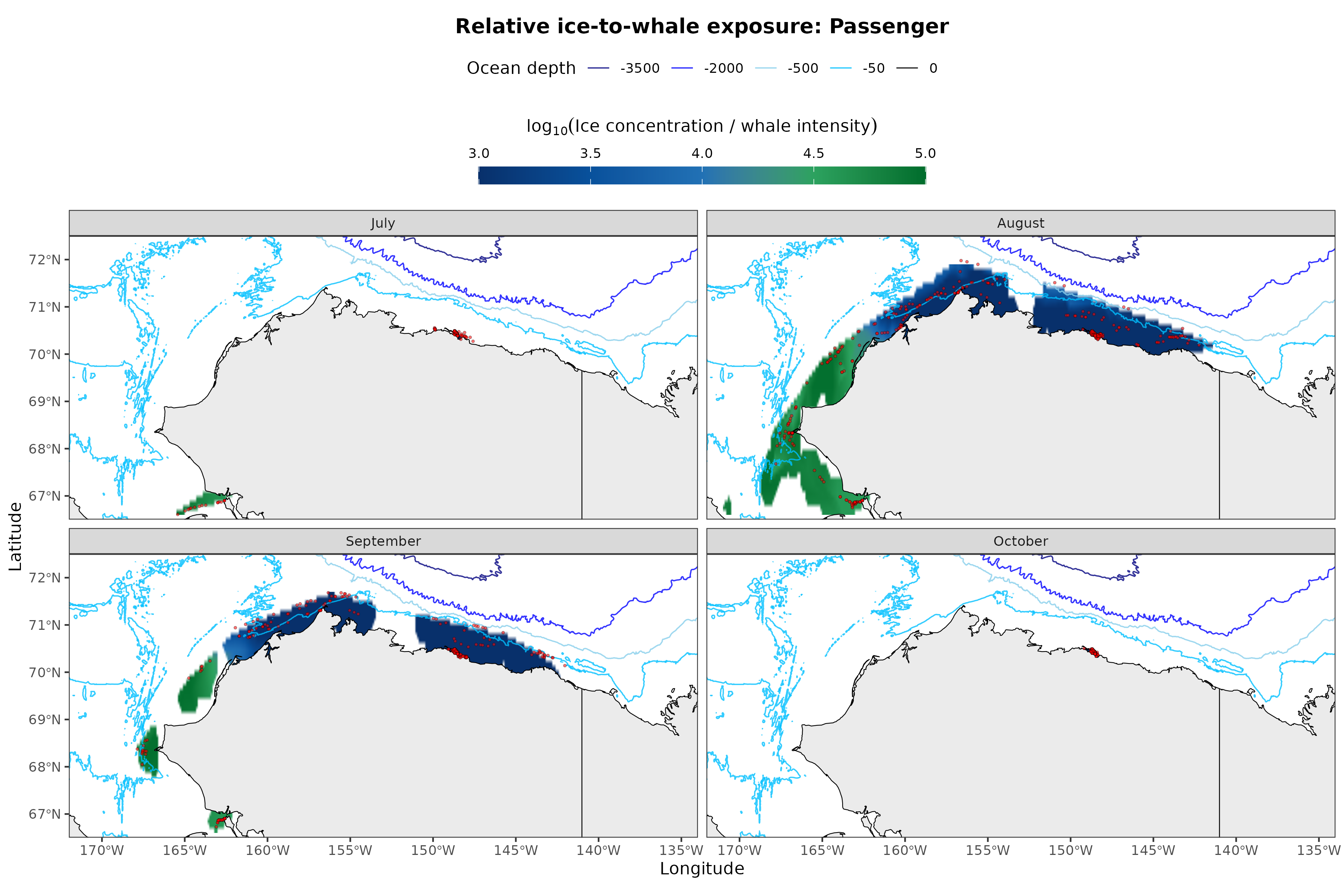}
}

\vspace{0.4cm}

\subfloat[Pilot Vessel]{
\includegraphics[width=0.95\textwidth]{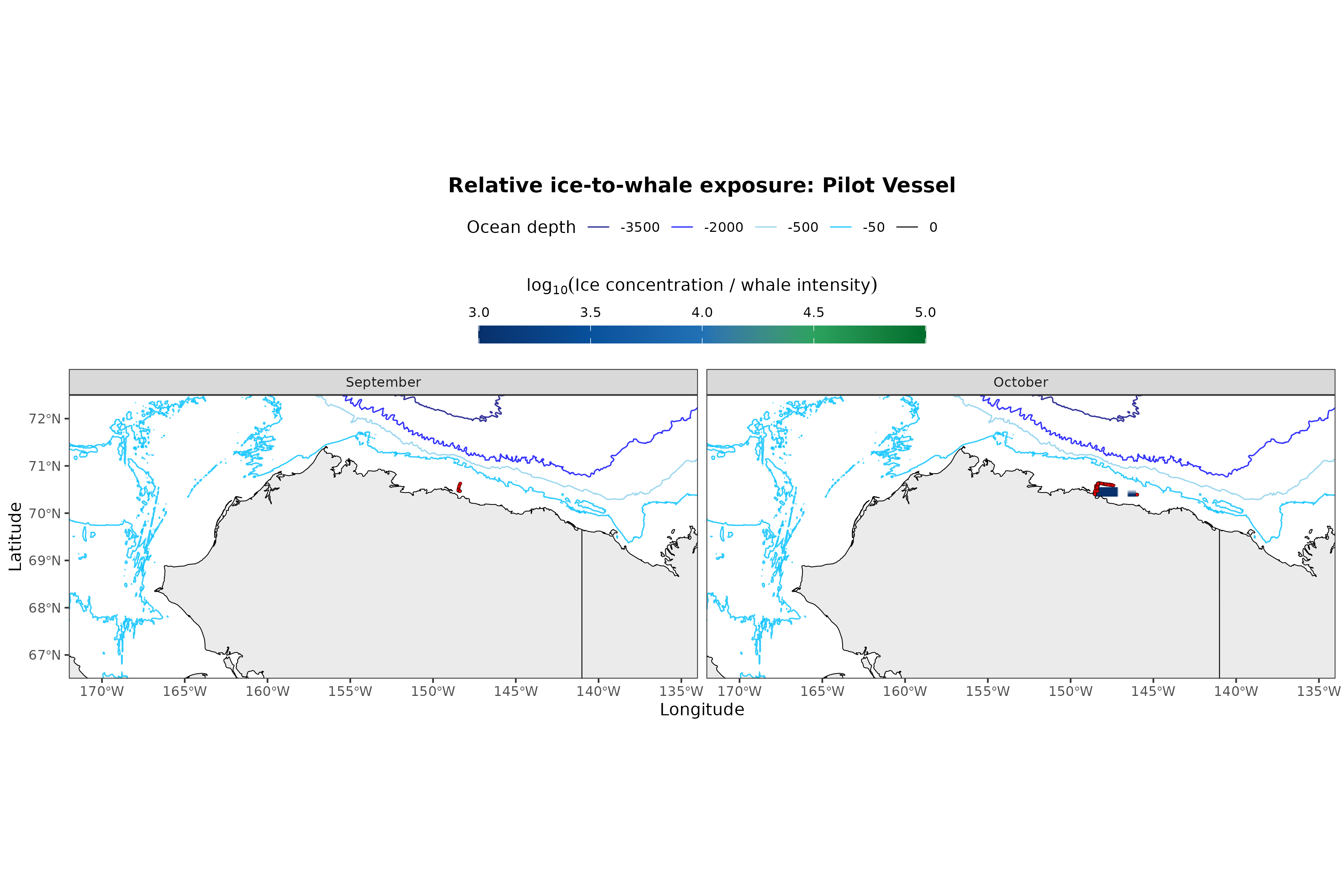}
}

\caption{Monthly spatial distribution of the log ratio of ice concentration to whale intensity.}
\label{fig:appendix_group_passenger_pilot}
\end{figure}

\begin{figure}[htbp]
\centering
\subfloat[Pleasure Craft]{
\includegraphics[width=0.95\textwidth]{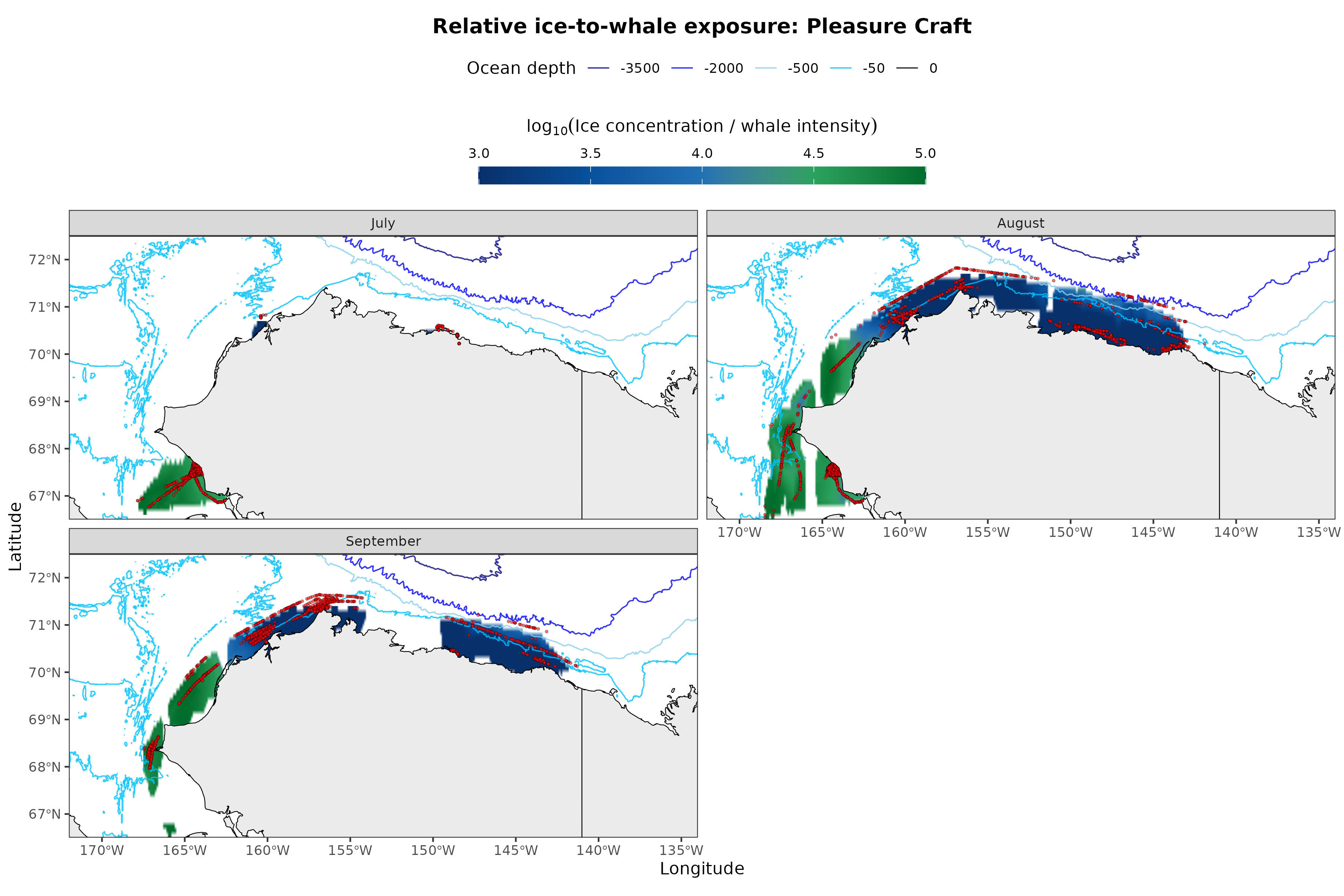}
}

\vspace{0.4cm}

\subfloat[Reserved]{
\includegraphics[width=0.95\textwidth]{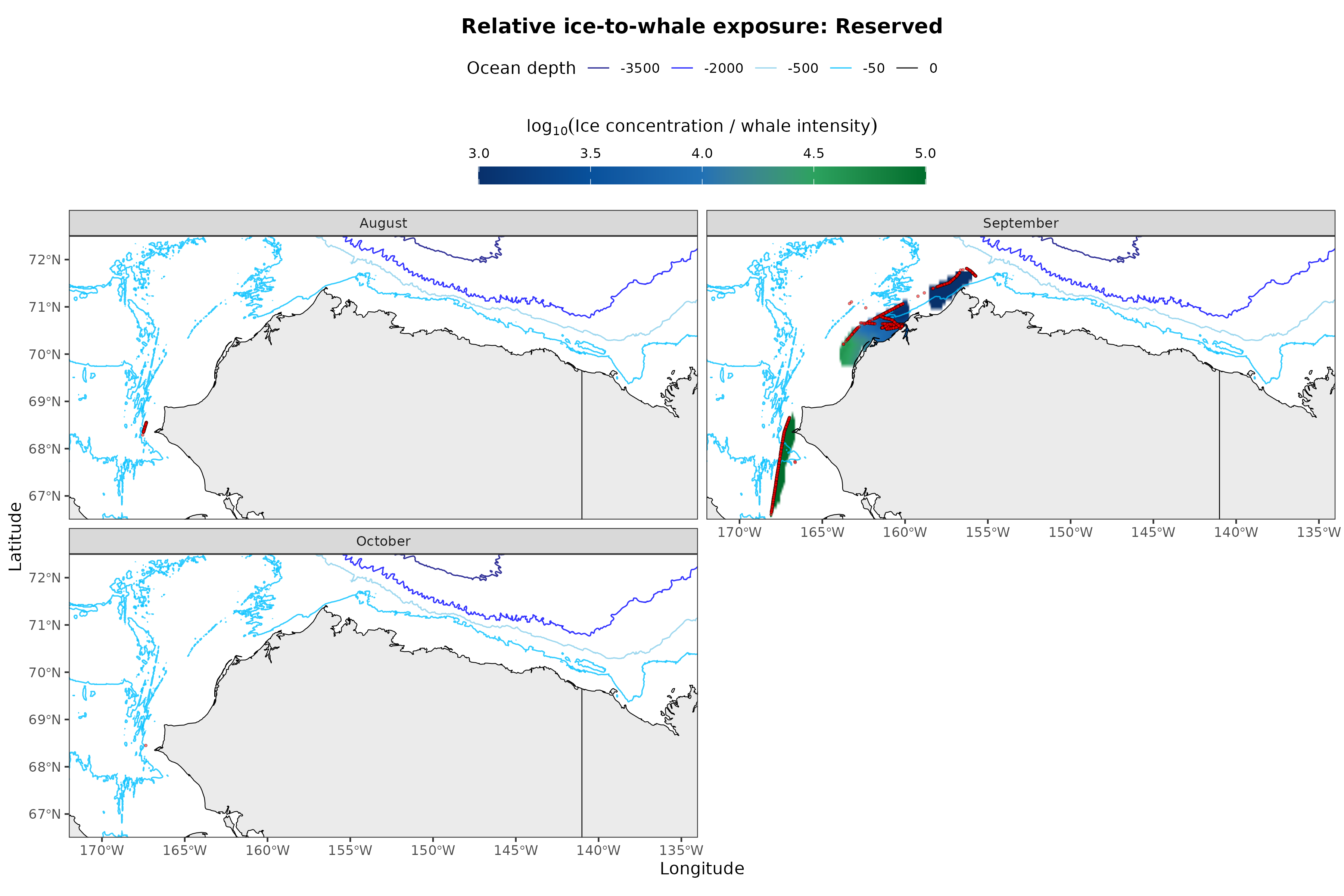}
}

\caption{Monthly spatial distribution of the log ratio of ice concentration to whale intensity.}
\label{fig:appendix_group_pleasure_reserved}
\end{figure}

\begin{figure}[htbp]
\centering
\subfloat[SAR]{
\includegraphics[width=0.95\textwidth]{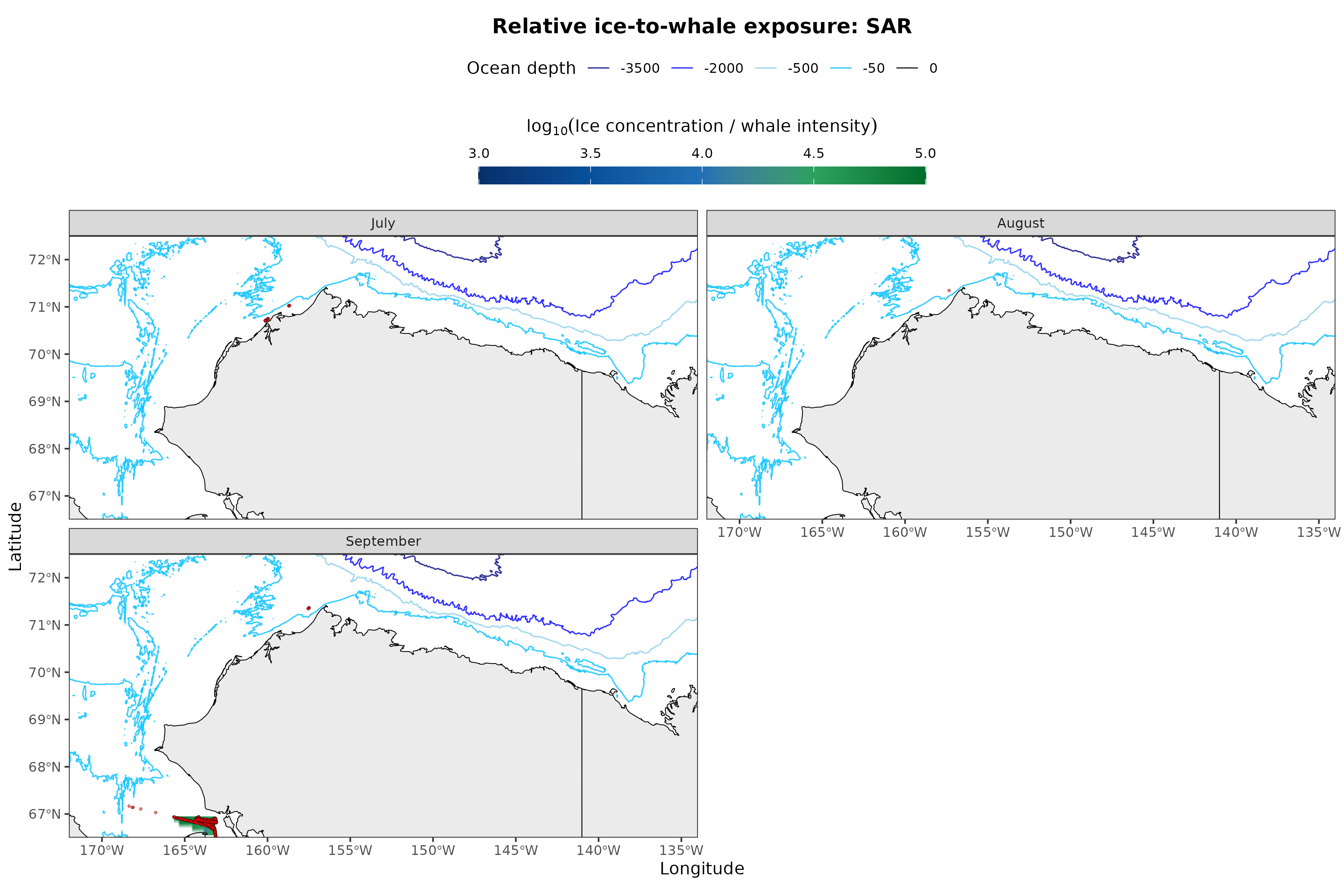}
}

\vspace{0.4cm}

\subfloat[Service Ship]{
\includegraphics[width=0.95\textwidth]{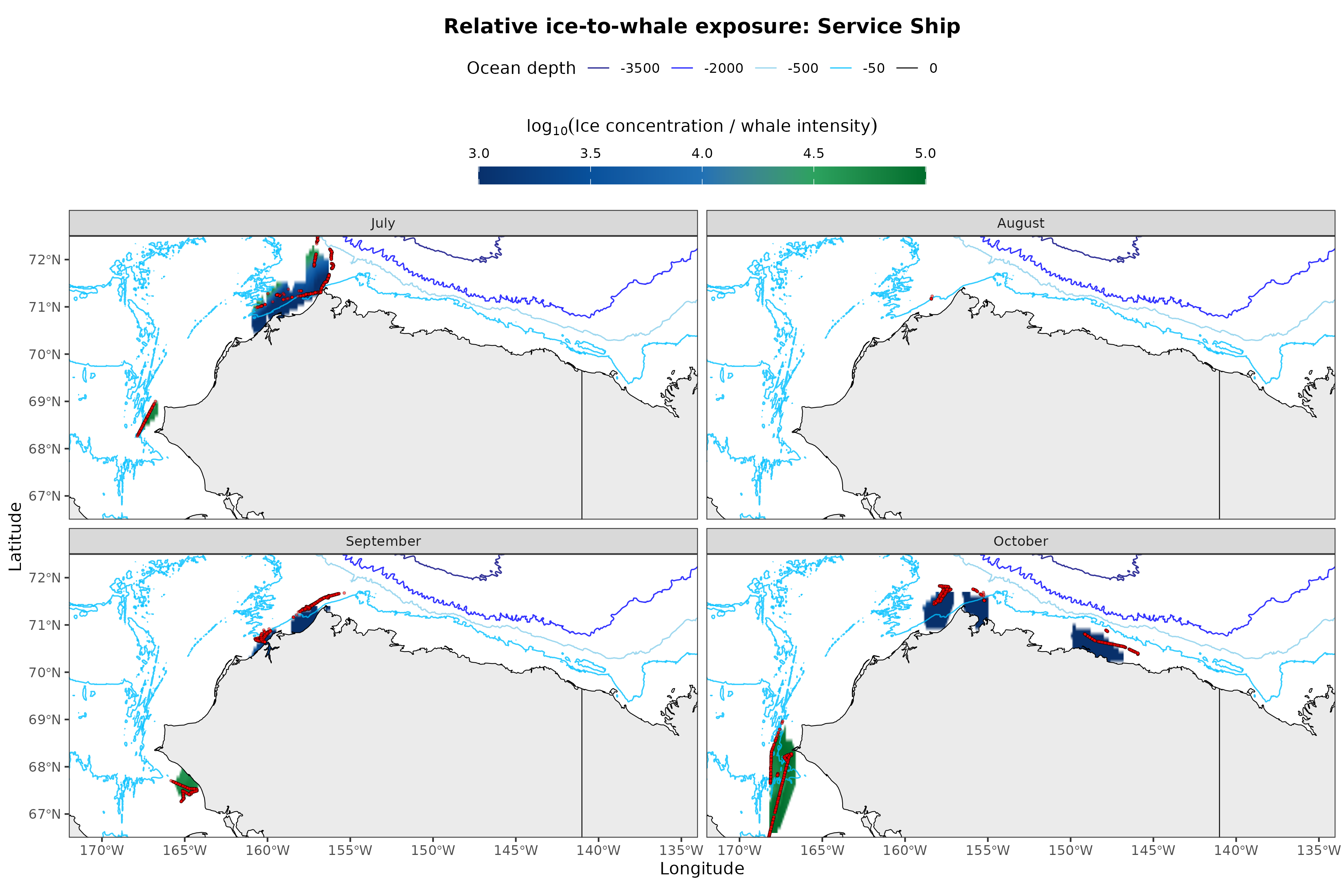}
}

\caption{Monthly spatial distribution of the log ratio of ice concentration to whale intensity.}
\label{fig:appendix_group_sar_service}
\end{figure}

\begin{figure}[htbp]
\centering
\subfloat[Tanker]{
\includegraphics[width=0.95\textwidth]{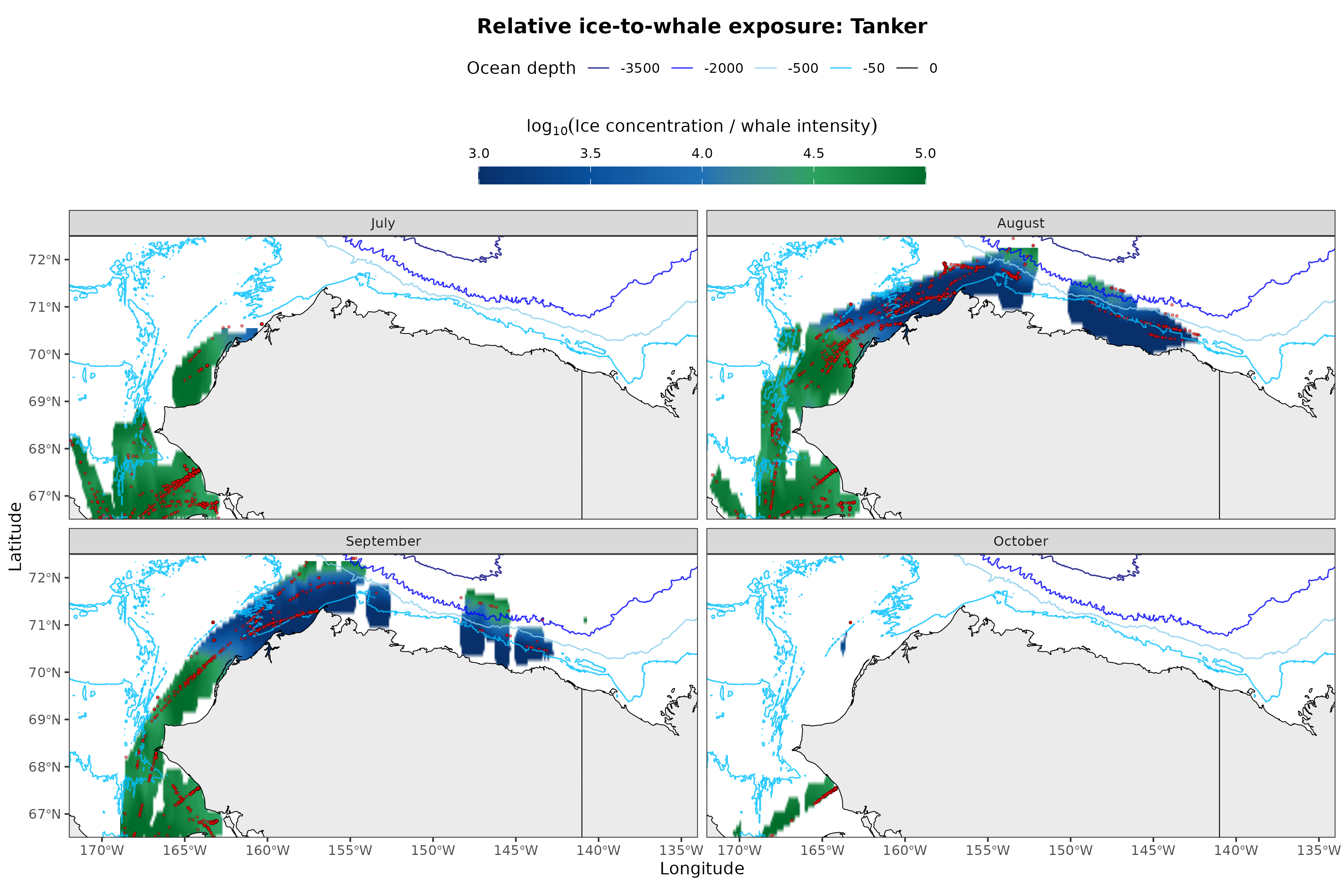}
}

\vspace{0.4cm}

\subfloat[Unspecified]{
\includegraphics[width=0.95\textwidth]{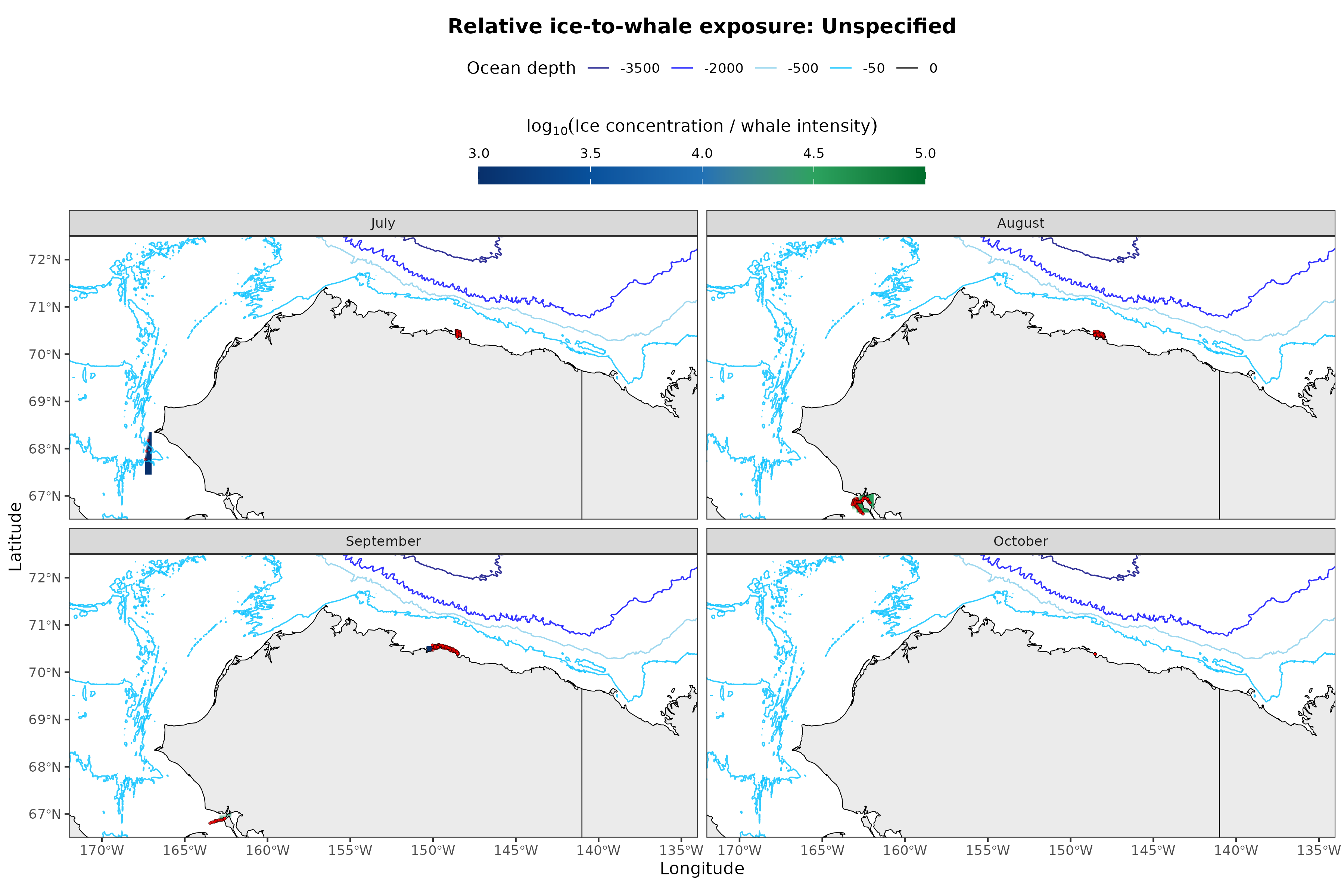}
}

\caption{Monthly spatial distribution of the log ratio of ice concentration to whale intensity.}
\label{fig:appendix_group_tanker_unspecified}
\end{figure}

\clearpage
\section{Status Ratio Plots}\label{secA2}

To examine how navigational status relate to environmental exposure, we construct monthly maps of the relative balance between ice and whale related risk across navigational status categories. For each status, observations are aggregated onto a spatial grid and the ratio of sea ice concentration to whale intensity is computed. We visualize the logarithm of this ratio, $\log_{10}\!\left(\frac{\text{ice concentration}}{\text{whale intensity}}\right)$, which provides a normalized measure of the dominant environmental constraint. Higher values indicate regions where ice exposure dominates, while lower values correspond to areas with relatively greater whale presence.

To ensure stability and interpretability, the ratio is averaged within grid cells and months, trimmed to remove extreme values, and interpolated only within regions supported by sufficient observations. Interpolation is further restricted using a convex hull around observed vessel locations for each status, preventing extrapolation into areas where the status is not observed. The resulting maps therefore capture the effective environmental conditions under which each operational state occurs. Overlaid points represent observed vessel positions, facilitating direct comparison between spatial exposure patterns and status specific activity.

\begin{figure}[htbp]
\centering
\subfloat[Status 0]{
\includegraphics[width=0.95\textwidth]{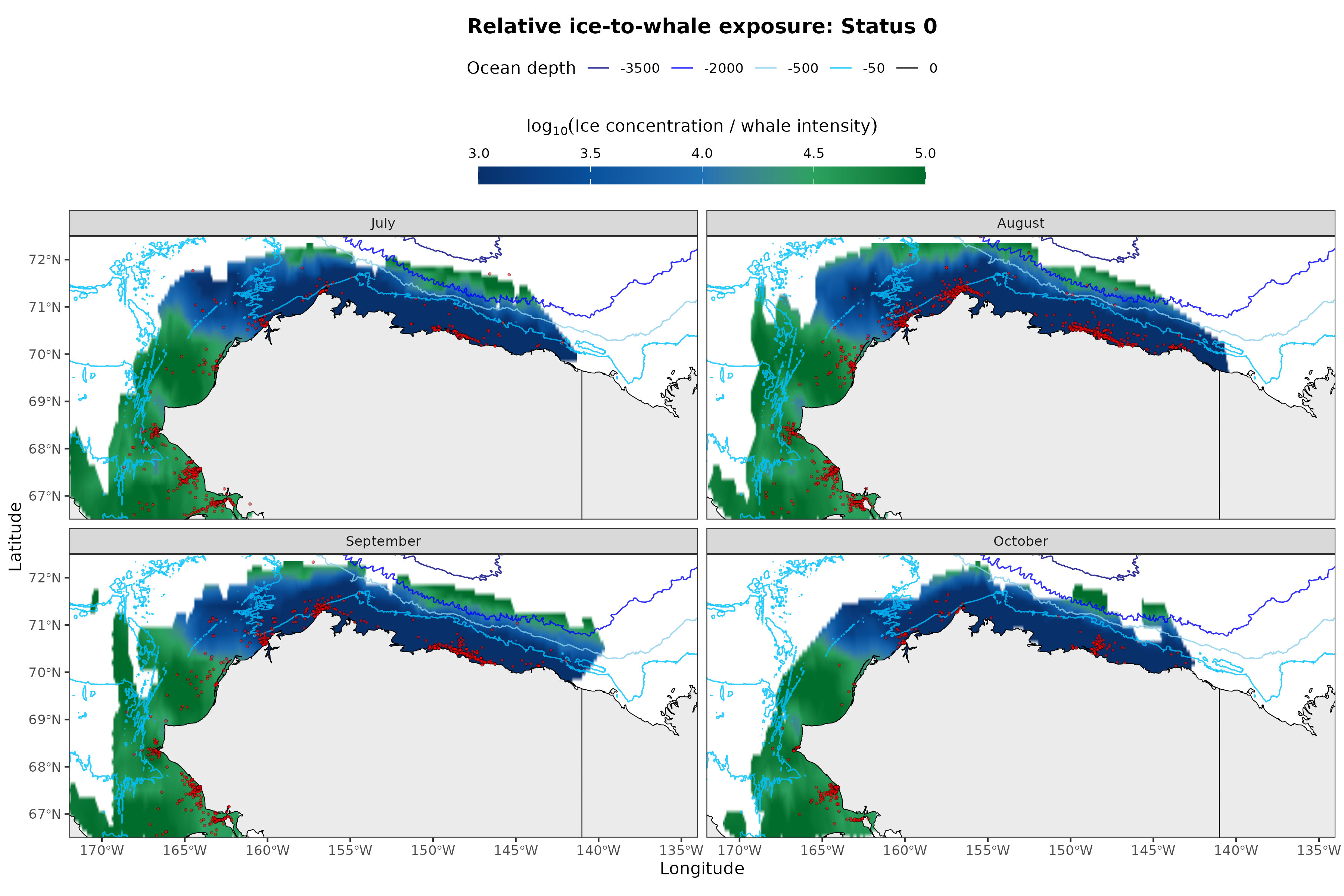}
}

\vspace{0.4cm}

\subfloat[Status 1]{
\includegraphics[width=0.95\textwidth]{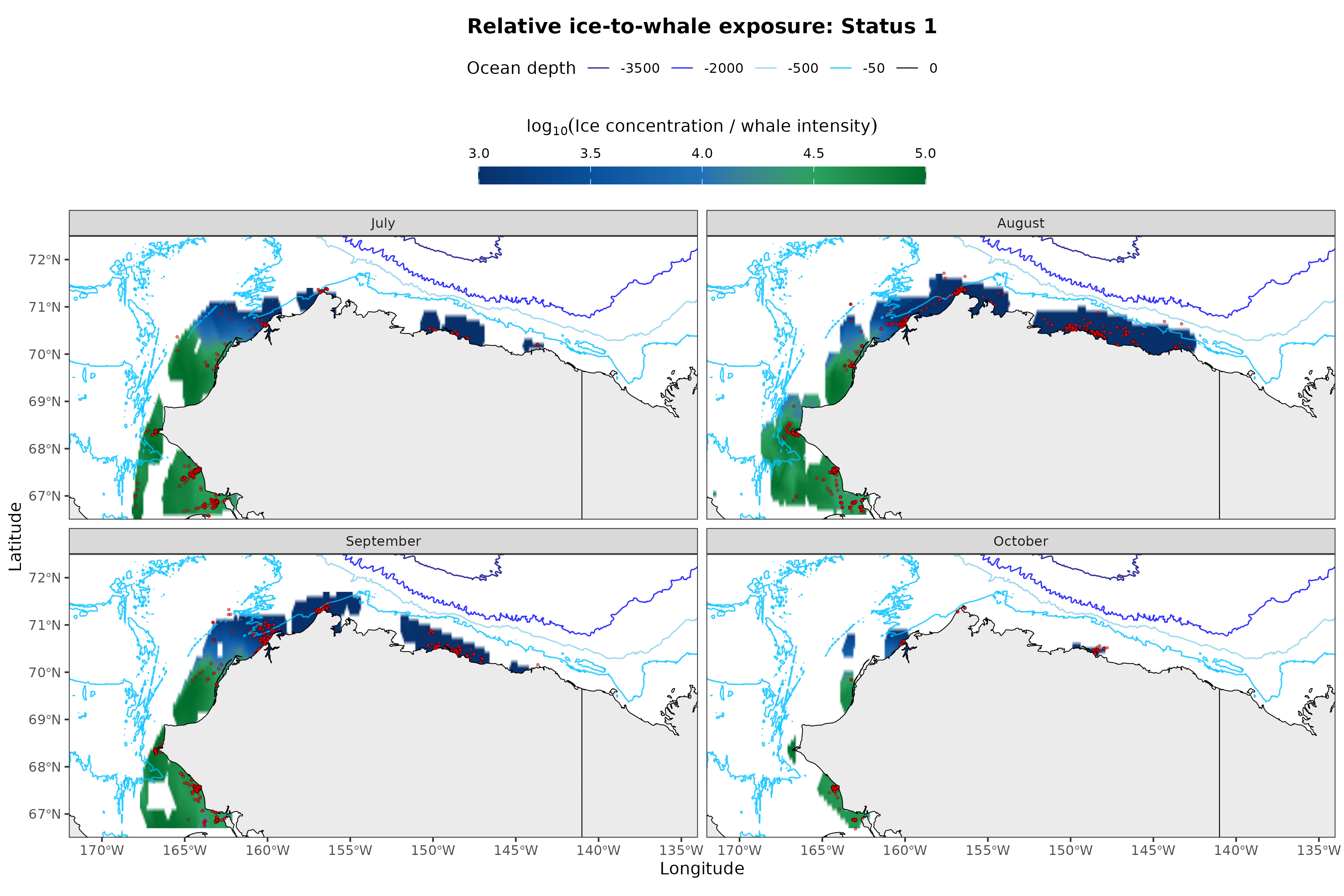}
}

\caption{Monthly spatial distribution of the log ratio of ice concentration to whale intensity by navigational status.}
\label{fig:appendix_status_0_1}
\end{figure}

\begin{figure}[htbp]
\centering
\subfloat[Status 2]{
\includegraphics[width=0.95\textwidth]{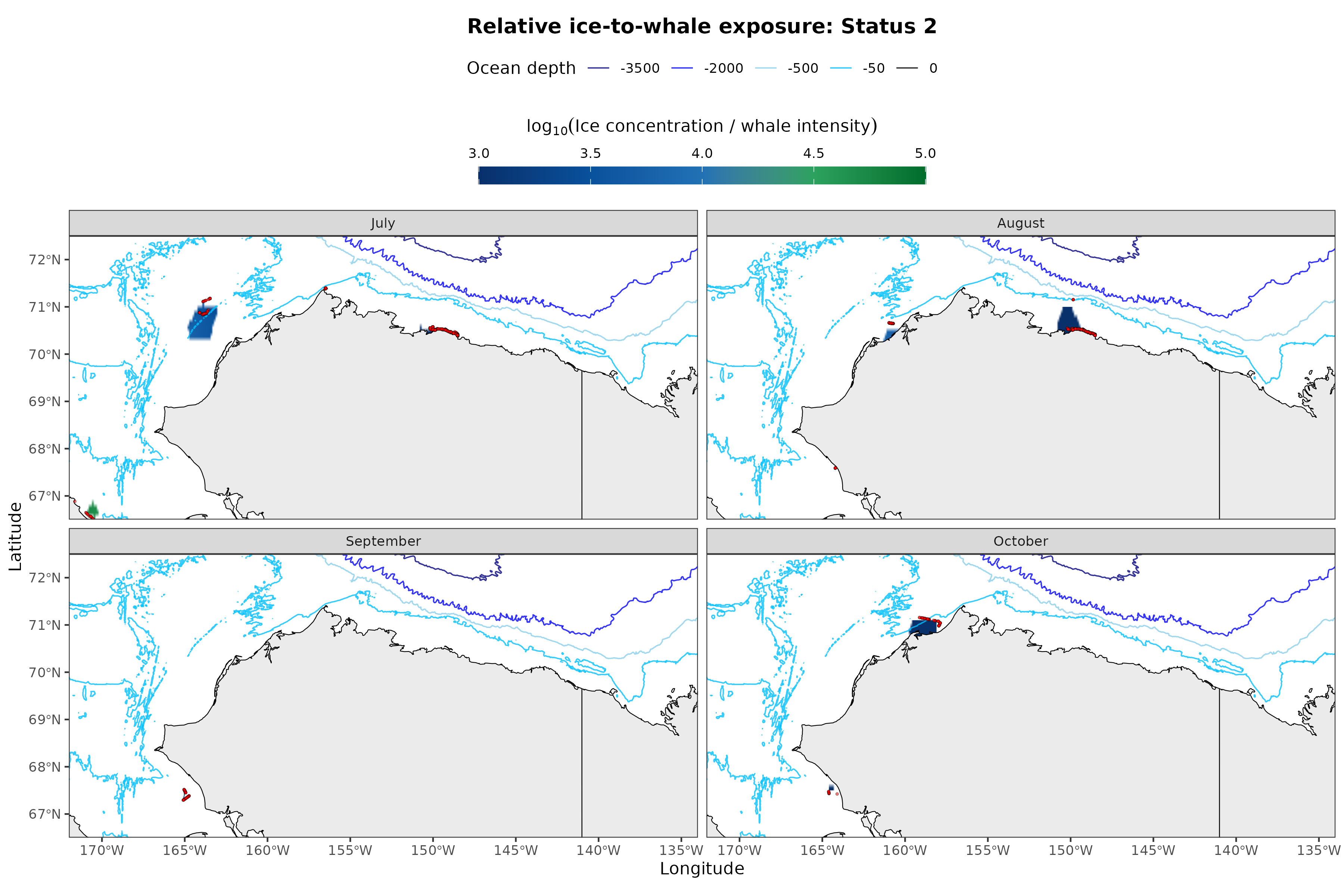}
}

\vspace{0.4cm}

\subfloat[Status 3]{
\includegraphics[width=0.95\textwidth]{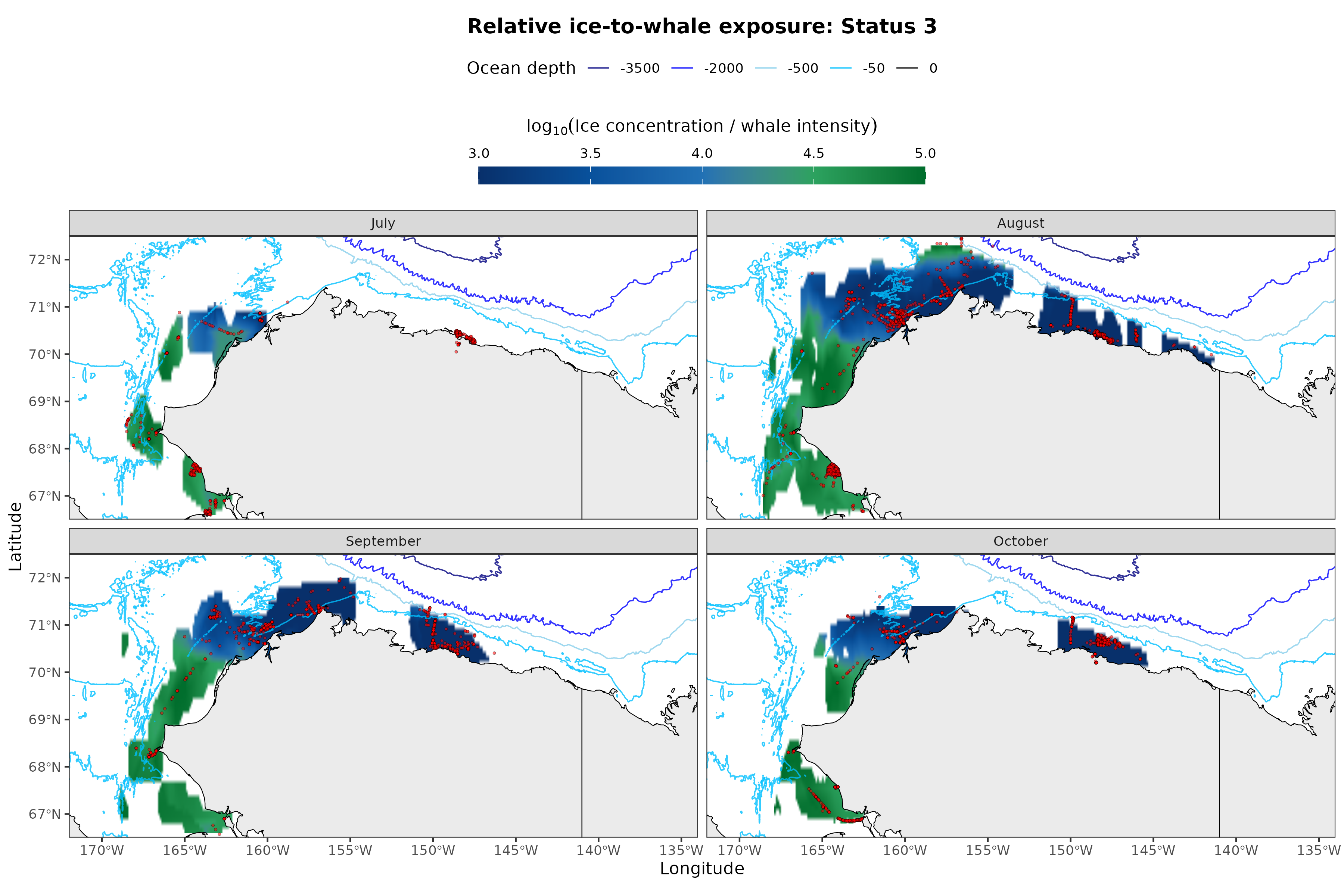}
}

\caption{Monthly spatial distribution of the log ratio of ice concentration to whale intensity by navigational status.}
\label{fig:appendix_status_2_3}
\end{figure}

\begin{figure}[htbp]
\centering
\includegraphics[width=0.95\textwidth]{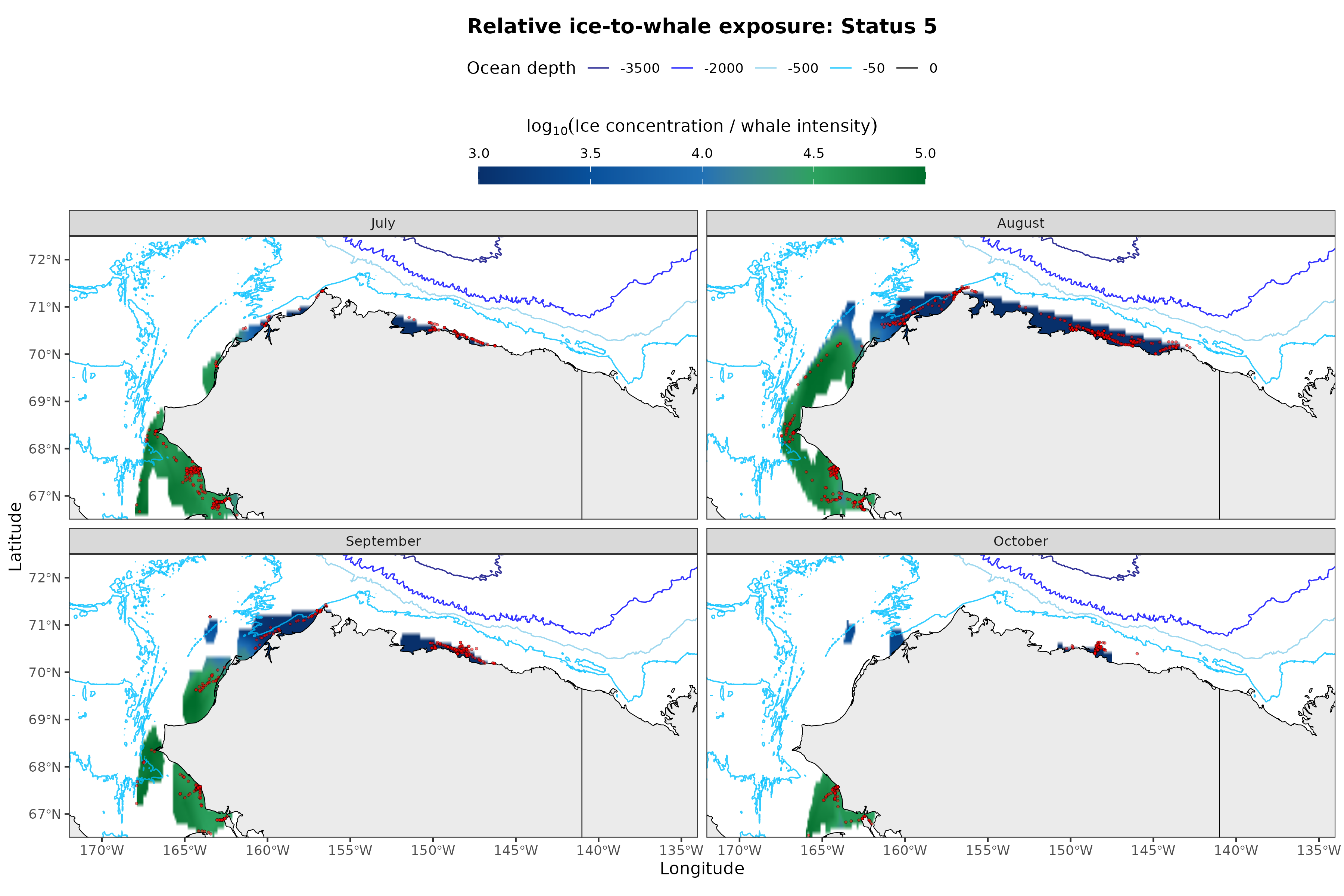}
\caption{Monthly spatial distribution of the log ratio of ice concentration to whale intensity for Status 5.}
\label{fig:appendix_status_5}
\end{figure}

\begin{figure}[htbp]
\centering
\includegraphics[width=0.95\textwidth]{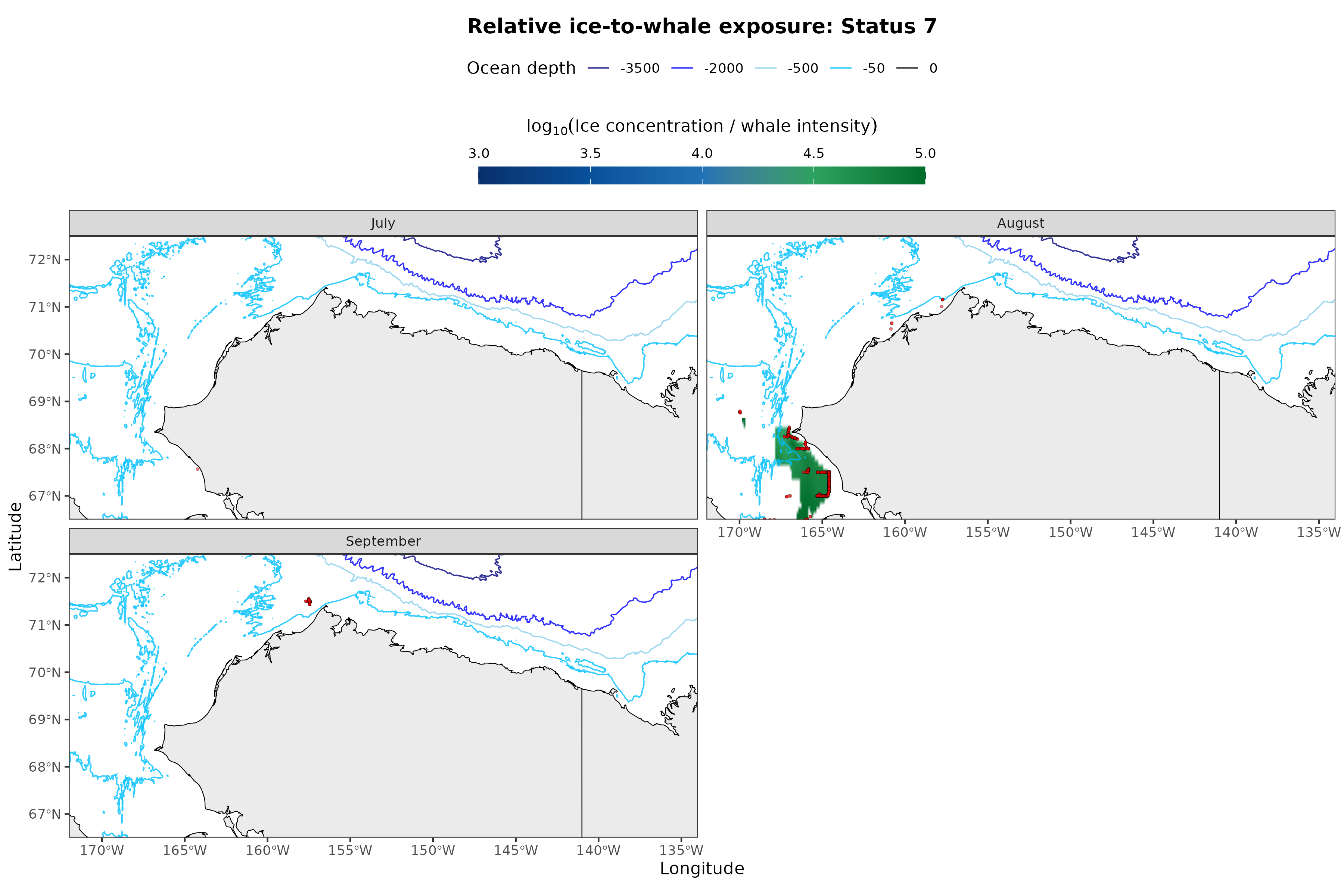}
\caption{Monthly spatial distribution of the log ratio of ice concentration to whale intensity for Status 7.}
\label{fig:appendix_status_7}
\end{figure}

\begin{figure}[htbp]
\centering
\includegraphics[width=0.95\textwidth]{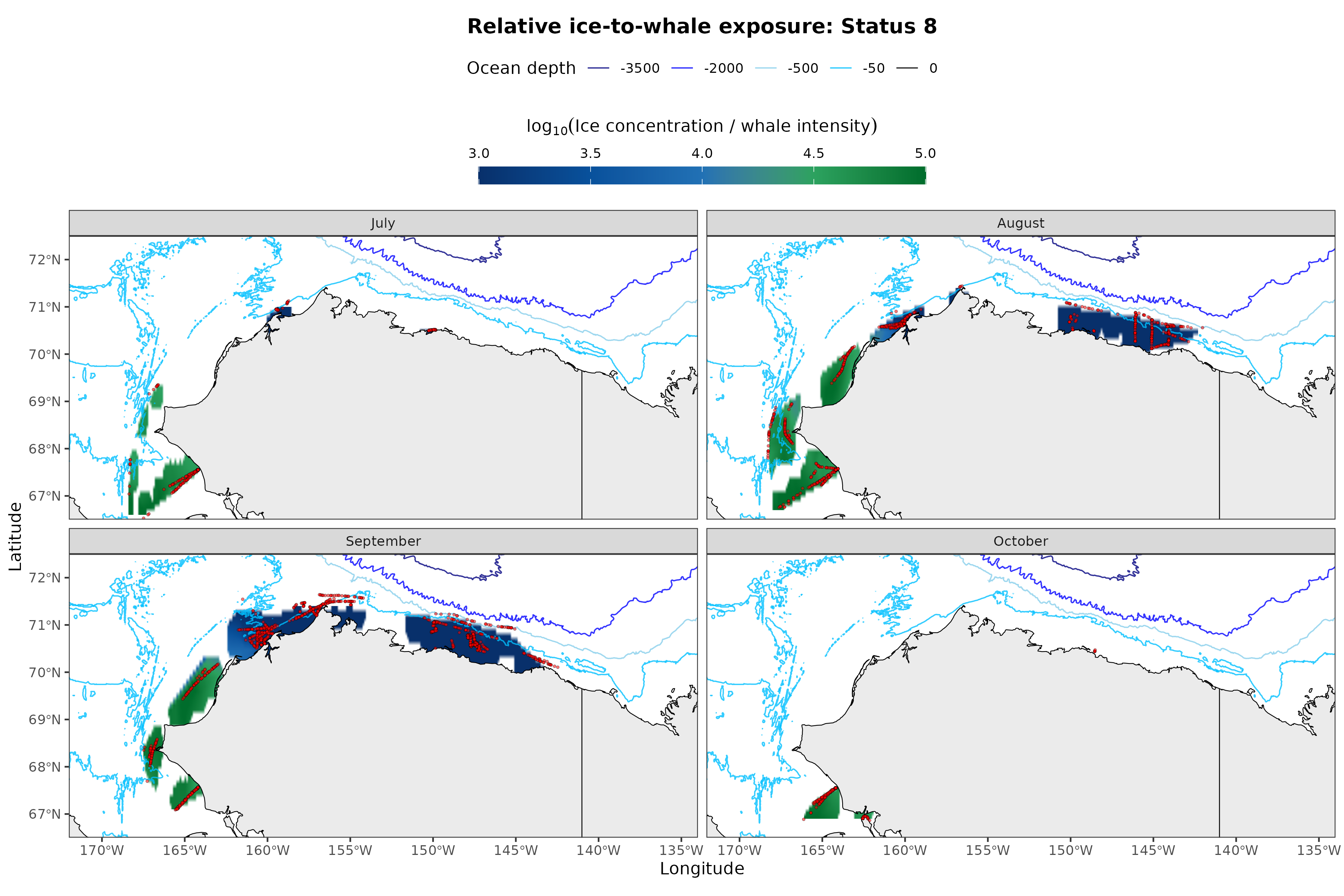}
\caption{Monthly spatial distribution of the log ratio of ice concentration to whale intensity for Status 8.}
\label{fig:appendix_status_8}
\end{figure}

\begin{figure}[htbp]
\centering
\includegraphics[width=0.95\textwidth]{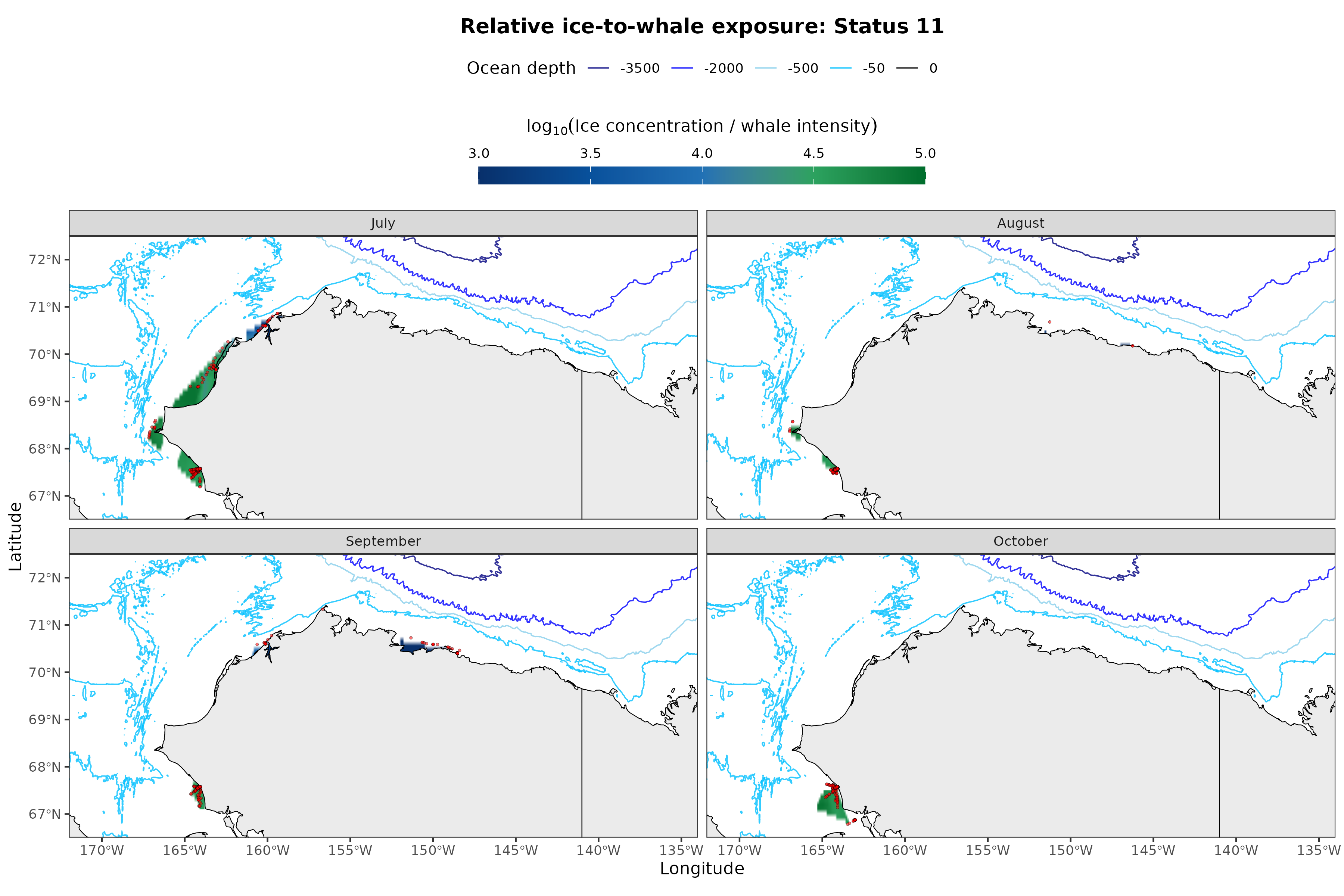}
\caption{Monthly spatial distribution of the log ratio of ice concentration to whale intensity for Status 11.}
\label{fig:appendix_status_11}
\end{figure}

\begin{figure}[htbp]
\centering
\includegraphics[width=0.95\textwidth]{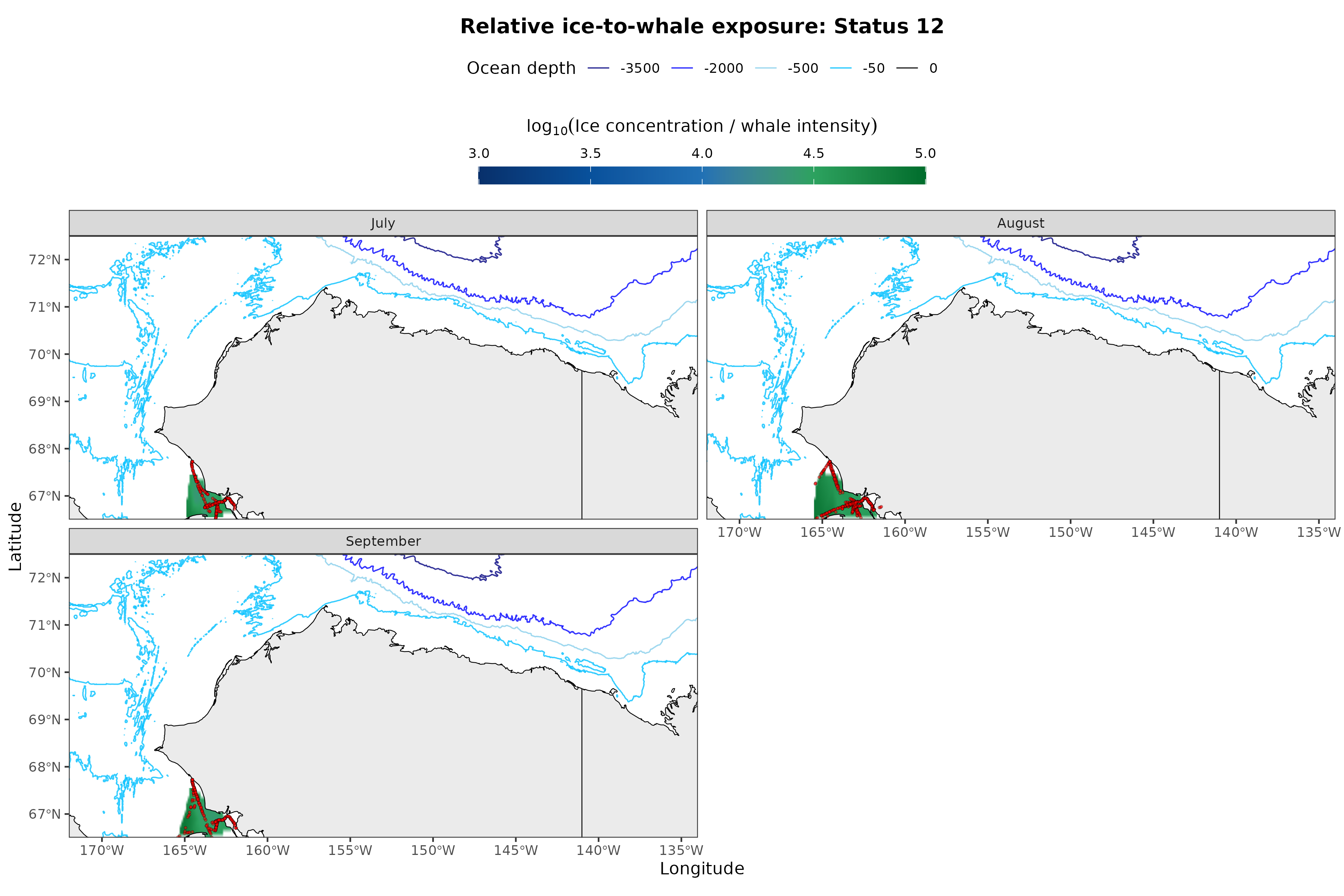}
\caption{Monthly spatial distribution of the log ratio of ice concentration to whale intensity for Status 12.}
\label{fig:appendix_status_12}
\end{figure}

\begin{figure}[htbp]
\centering
\includegraphics[width=0.95\textwidth]{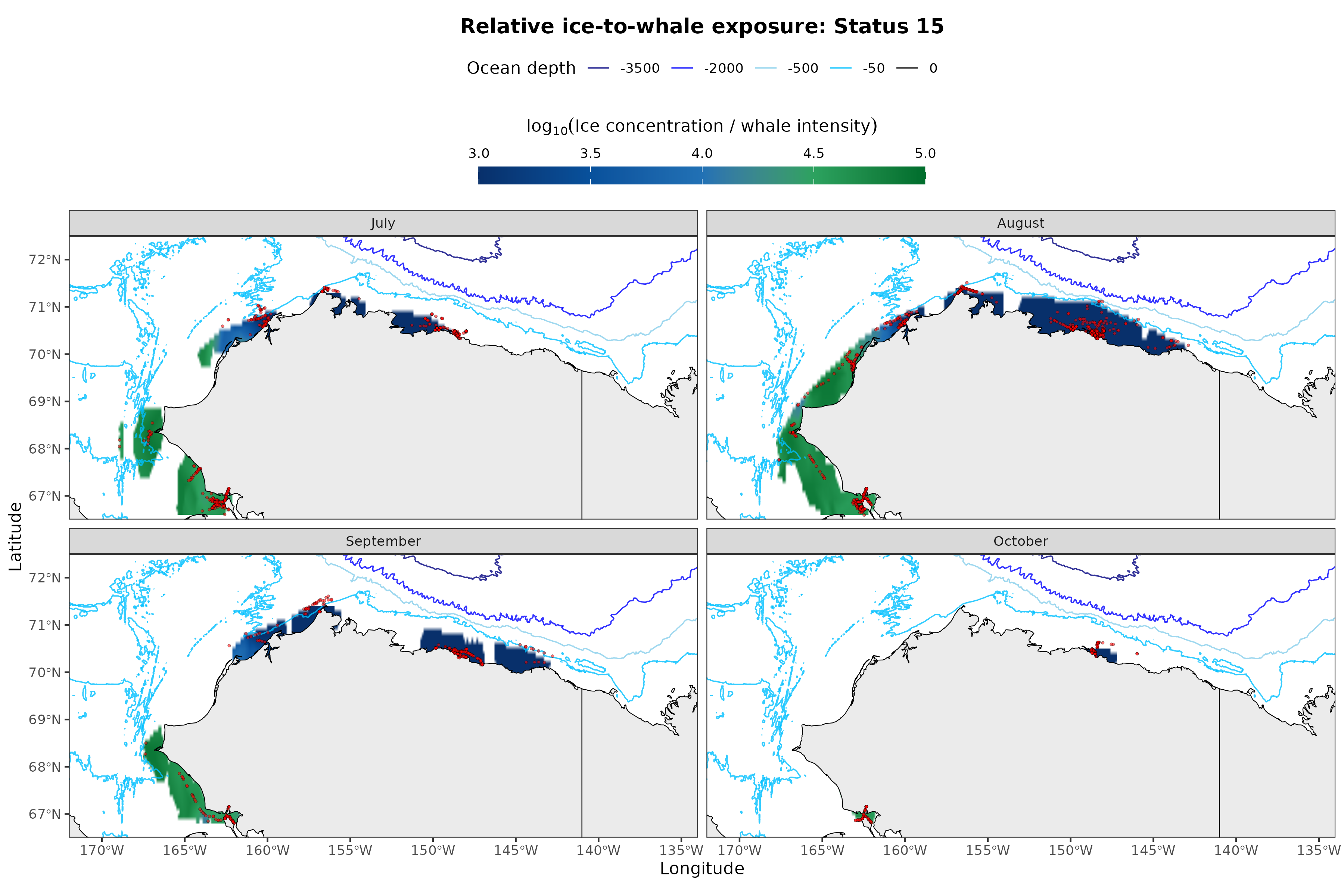}
\caption{Monthly spatial distribution of the log ratio of ice concentration to whale intensity for Status 15.}
\label{fig:appendix_status_15}
\end{figure}

\end{document}